\documentclass[preprint]{ptephy_v1}

\usepackage{graphicx, epsfig}

\newcommand{\tr}{\mbox{tr}}

\usepackage{color}
\usepackage{physics}
\usepackage{hyperref}
\usepackage{booktabs}

\newcommand{\lrpdv}{\overset{\substack{\text{$\leftrightarrow$} \\ \vspace{-2.1ex}}}{\partial}{}^{}}

\begin{document}

\title{
Gravitational waves from first-order phase transition in an electroweakly interacting vector dark matter model}

\author{Tomohiro~Abe}
\affil{
Department of Physics, Faculty of Science and Technology, Tokyo University of Science, Noda, Chiba 278-8510, Japan
\email{abe.tomohiro@rs.tus.ac.jp}}
\author[1,2]{Katsuya~Hashino}
\affil{
National Institute of Technology, Fukushima College, Nagao 30, Taira-Kamiarakawa, Iwaki, Fukushima 970–8034, Japan
\email{hashino@fukushima-nct.ac.jp}}

\begin{abstract}

We discuss gravitational waves in an electroweakly interacting vector dark matter model. 
In the model, the electroweak gauge symmetry is extended to
SU(2)$_0 \times$ SU(2)$_1 \times$SU(2)$_2 \times$ U(1)$_Y$ and spontaneously broken into SU(2)$_L \times$ U(1)$_Y$ at TeV scale.
The model has an exchange symmetry between SU(2)$_0$ and SU(2)$_2$. This symmetry stabilizes some massive vector bosons associated with the spontaneous symmetry breaking described above, and an electrically neutral one is a dark matter candidate.
In the previous study, it was found that the gauge couplings of SU(2)$_0$ and SU(2)$_1$ are relatively large to explain the measured value of the dark matter energy density via the freeze-out mechanism. 
With the large gauge couplings, the gauge bosons potentially have a sizable effect on the scalar potential. 
In this paper, we focus on the phase transition of SU(2)$_0 \times$ SU(2)$_1 \times$ SU(2)$_2 \to$ SU(2)$_L$.
We calculate the effective potential at finite temperature and find that the phase transition is first-order and strong in a wide range of the parameter space. The strong first-order phase transition generates gravitational waves. We calculate the gravitational wave spectrum and find that it is possible to detect the gravitational waves predicted in the model by future space-based gravitational wave interferometers.
We explore the regions of the parameter space probed by the gravitational wave detection. We find that the gravitational wave detection can probe the region where the mass of $h'$, a CP-even scalar in the model, is a few TeV.

\end{abstract}

\maketitle

\section{Introduction}
\label{sec:intro}

Many astrophysical observations show the existence of dark matter (DM). 
DM constitutes approximately 26\% of the energy in the universe~\cite{Planck:2018vyg}. 
However, the nature of DM remains unclear. 
Models beyond the standard model (SM) of particle physics often predict DM candidates.
In many particle DM models, the freeze-out mechanism~\cite{Lee:1977ua} is used to explain the measured value of the DM energy density.
The mechanism requires a pair of DM particles to annihilate into other particles in the thermal bath in the early Universe.
The canonical value of the annihilation cross section, which can explain the measured value of the DM energy density, is $\expval{\sigma v} \simeq 3 \times 10^{-26}$ cm$^3$~s$^{-1} \simeq 1$~pb $c$. This value is of the same order as the cross section of processes by the electroweak interaction and implies that DM particles interact with the SM particles via the electroweak interaction. An example of such DM is the wino DM, which is an SU(2)$_L$ triplet spin-$\frac{1}{2}$ Majorana fermion, and the mass prediction of the thermally produced wino DM is approximately 3~TeV~\cite{Hisano:2006nn, Cirelli:2007xd}.

The electroweakly interacting vector DM model proposed in~\cite{Abe:2020mph} is one of the attractive DM models. In the model, the electroweak gauge symmetry SU(2)$_L \times$ U(1)$_Y$ is extended into SU(2)$_0 \times$SU(2)$_1 \times$SU(2)$_2 \times$U(1)$_Y$. The SU(2)$_0 \times$SU(2)$_1 \times$SU(2)$_2$ gauge symmetry is spontaneously broken into SU(2)$_L$ by vacuum expectation values of two scalar fields $\Phi_1$ and $\Phi_2$. An exchange symmetry SU(2)$_0 \leftrightarrow$SU(2)$_2$ is imposed, and it stabilizes linear combinations of gauge fields $V^a_\mu \equiv (W_{0\mu}^a - W_{2\mu}^a)/\sqrt{2}$, where $W_{j\mu}^a$ is the gauge field of SU(2)$_j$ with $j = 0,1,2$. After the electroweak symmetry breaking SU(2)$_L \times$U(1)$_Y \to $U(1)$_\text{em}$, quantum corrections make charged component of $V^a$, which is denoted as $V^\pm$, slightly heavier than the neutral component $V^0$. Consequently, $V^0$ is a DM candidate. Note that $V^a$ arises from the SU(2)$_0$ and SU(2)$_2$ gauge fields, and the SU(2)$_L$ gauge field are linear combination of the SU(2)$_0$, SU(2)$_1$, and SU(2)$_2$ gauge fields. Hence, $V^a$ directly couples to the SU(2)$_L$ gauge bosons and have the electroweak interaction. 
Therefore, the model predicts electroweakly interacting vector DM.

Through the electroweak interaction, $V^a$ is in equilibrium with the SM particles in the early universe, and the DM abundance is determined by the freeze-out mechanism. In addition to $V^a$ and the SM gauge bosons, the model predicts extra heavy gauge bosons, denoted by $W'^\pm$ and $Z'$. They are approximately SU(2)$_L$ triplet and have common mass, $m_{W'} \simeq m_{Z'}$. Annihilation of a pair of $V^0$ depend on $m_{Z'}$. For example, if $m_{Z'} \lesssim 2 m_V - m_W$, $V^0 V^0 \to W'^- W^+$ is open, but otherwise it is kinematically forbidden. Hence, the value of $m_{Z'}$ affects the DM relic abundance. The mass of the vector DM that reproduces the measured value of the DM energy density varies from $3$~TeV to $19$~TeV~\cite{Abe:2020mph}.

Other than relic abundance, direct detection of $V^0$, $W'/Z'$ search in collider experiments, and indirect detection of DM were studied in~\cite{Abe:2020mph, Abe:2021mry}. However, these observables are insufficient to probe all the regions of the model parameter space. Other observables are necessary to test the model comprehensively.

In this work, we focus on gravitational waves (GWs), aiming to extend the region of the parameter space where we can test the electroweakly interacting DM model with future/current experiments. Because the gauge couplings $g_0$ and $g_1$ are $\sim 1$ to explain the relic abundance~\cite{Abe:2020mph}, the scalar potential is modified by the gauge boson contributions at the loop level. As a result, the scalar potential can experience a first-order phase transition in the early universe. It is well-known that first-order phase transitions generate GWs~\cite{Grojean:2006bp}. To observe the GW spectra, future space-based GW interferometers, such as LISA~\cite{LISA}, DECIGO~\cite{DECIGO}, and BBO~\cite{BBO}, can be used; thus, DM models with a first-order phase transition can be tested through GW observational experiments~\cite{Schwaller:2015tja,Chala:2016ykx,Baldes:2017rcu,Chao:2017vrq,Beniwal:2017eik,Addazi:2017gpt,Tsumura:2017knk,Huang:2017rzf,Huang:2017kzu,Hektor:2018esx,Hashino:2018zsi,Baldes:2018emh,Madge:2018gfl,Beniwal:2018hyi,Bian:2018mkl,Bai:2018dxf,Bian:2018bxr,Shajiee:2018jdq,Mohamadnejad:2019vzg,Bertone:2019irm,Kannike:2019mzk,Paul:2019pgt,Croon:2019rqu,Hall:2019ank,Chen:2019ebq,Hall:2019rld,Barman:2019oda,Chiang:2019oms,Borah:2020wut,Kang:2020jeg,Pandey:2020hoq,Hong:2020est,Alanne:2020jwx,Bhoonah:2020oov,Han:2020ekm,Wang:2020wrk,Ghosh:2020ipy,Huang:2020crf,Deng:2020dnf,Chao:2020adk,Azatov:2021ifm,Zhang:2021alu,Davoudiasl:2021ijv,Reichert:2021cvs,Mohamadnejad:2021tke,Bian:2021dmp,Costa:2022oaa,Liu:2022jdq,Shibuya:2022xkj,Costa:2022lpy,Kierkla:2022odc,Morgante:2022zvc,Chakrabarty:2022yzp,Arcadi:2022lpp,Frandsen:2022klh}. We investigate the GWs generated by the phase transition from SU(2)$_0 \times$SU(2)$_1 \times$SU(2)$_2$ to SU(2)$_L$ in the electroweakly interacting vector DM model.
As discussed below, it allows us to explore the region of the parameter space where other observables, such as the $W'$ search, cannot probe.

The rest of this paper is organized as follows.
In Section~\ref{sec:model}, we briefly introduce the vector DM model with ${\rm SU}(2)_0\times{\rm SU}(2)_1\times{\rm SU}(2)_2\times {\rm U}(1)_Y$ gauge symmetry. 
In Section~\ref{sec:potential}, we show the effective potential with finite-temperature effects for this model. 
We clarify the parameter region with a first-order phase transition, which can produce a detectable GW spectrum. 
The formula for the GW spectrum from the first-order phase transition is presented in Section~\ref{sec:GW}.
We discuss in Section~\ref{sec:numerical} the testability of the model at the GW observation experiments, such as the LISA, DECIGO, and BBO experiments. 
Finally, Section~\ref{sec:conclusion} presents the conclusions of this study. 
In appendix~\ref{app:Veff-at-T=0}, we give the explict expression of the effective potential at the zero temperature.

\section{The model}
\label{sec:model}

In this section, we introduce the electroweakly interacting vector DM model proposed in~\cite{Abe:2020mph}.

The model exhibits gauge symmetry, described as $\text{SU(3)}_c \times \text{SU(2)}_0\times \text{SU(2)}_1 \times \text{SU(2)}_2 \times \text{U(1)}_Y$. Here, $\text{SU(3)}_c$ and $\text{U(1)}_Y$ correspond to the gauge symmetries governing the quantum chromodynamics (QCD) and hypercharge, respectively.
Because the QCD sector is the same as that in the SM, we focus on the electroweak sector, denoted as  $\text{SU(2)}_0 \times \text{SU(2)}_1 \times \text{SU(2)}_2 \times \text{U(1)}_Y$.
We use the notation $W^{a}_{j \mu}$ and $B_\mu$ for gauge bosons associated with $\text{SU(2)}_j$ and $\text{U(1)}_Y$, respectively. Here, $j$ can take values of 0, 1, or 2, and $a$ can take values of 1, 2, or 3. 
$g_j$ and $g'$ are the gauge couplings for $\text{SU(2)}_j$ and $\text{U(1)}_Y$, respectively.

We introduce two scalar fields, $\Phi_1$ and $\Phi_2$, expressed in two-by-two matrices. They transform under gauge transformation as
\begin{align}
\Phi_1 \to U_0 \Phi_1 U_1^\dagger,\
\Phi_2 \to U_2 \Phi_1 U_1^\dagger,
\end{align}
where $U_0, U_1$, and $U_2$ represent two-by-two unitary matrices for $\text{SU(2)}_0, \text{SU(2)}_1$, and $\text{SU(2)}_2$, respectively.
In addition, we impose the following conditions for $\Phi_j$ to reduce their degrees of freedom,
\begin{equation}
\Phi_j = - \epsilon \Phi_j^* \epsilon,\ \ \ \text{where } \epsilon = \mqty(0 & 1 \\ -1 & 0).
\end{equation}
Hence, each $\Phi_j$ consists of four real scalar fields.

All other fields remain identical to those in the SM, except that they are charged under $\text{SU(2)}_1$ instead of $\text{SU(2)}_L$. The charge assignments for the matter fields are summarized in Table~\ref{tab:charge}. 
\begin{table}[tb]
\centering
 \begin{tabular}{cc|ccccc}\hline
  field    & spin           & SU(3)$_c$ & SU(2)$_0$ & SU(2)$_1$ & SU(2)$_2$ & U(1)$_Y$ \\ \hline \hline
  $q_L$    & $\frac{1}{2}$  &   3   &   1       &    2      &     1     &   $\frac{1}{6}$ \\
  $u_R$    & $\frac{1}{2}$  &   3   &   1       &    1      &     1     &   $\frac{2}{3}$ \\
  $d_R$    & $\frac{1}{2}$  &   3   &   1       &    1      &     1     &  -$\frac{1}{3}$ \\
  $\ell_L$ & $\frac{1}{2}$  &   1   &   1       &    2      &     1     &  -$\frac{1}{2}$ \\
  $e_R$    & $\frac{1}{2}$  &   1   &   1       &    1      &     1     &  -1 \\ \hline
  $H$      & 0              &   1   &   1       &    2      &     1     &   $\frac{1}{2}$ \\
  $\Phi_1$ & 0              &   1   &   2       &    2      &     1     &   0 \\
  $\Phi_2$ & 0              &   1   &   1       &    2      &     2     &   0 \\ \hline \hline
 \end{tabular}
 \caption{Charge assignment under the SU(3)$_C\times$SU(2)$_0\times$SU(2)$_1\times$SU(2)$_2$ $\times$U(1)$_Y$ gauge symmetry of the model.}
 \label{tab:charge}
\end{table}

In addition to gauge symmetry, this model exhibits exchange symmetry. 
The Lagrangian is invariant under the following field transformations:
\begin{align}
W^a_{0\mu} \to W^a_{2\mu}, \
W^a_{2\mu} \to W^a_{0\mu}, \
\Phi_{1} \rightarrow \Phi_{2}, \
\Phi_{2} \rightarrow \Phi_{1},
\label{eq:exchange-sym}
\end{align}
whereas all other fields remain unchanged. 
This symmetry is equivalent to the exchange between $\text{SU(2)}_0$ and $\text{SU(2)}_2$, implying that the gauge couplings of $\text{SU(2)}_0$ and $\text{SU(2)}_2$ must be identical.
It is important to note that under this symmetry, 
$\frac{W^{a}_{0\mu} - W^{a}_{2\mu}}{\sqrt{2}}$ and $\frac{\Phi_{1} - \Phi_2}{\sqrt{2}}$ change sign, whereas $\frac{W^{a}_{0\mu} + W^{a}_{2\mu}}{\sqrt{2}}$, $\frac{\Phi_{1} + \Phi_2}{\sqrt{2}}$, and the other fields remain unchanged.
Therefore, the symmetry described in Eq.~\eqref{eq:exchange-sym} is equivalent to the $Z_2$ symmetry commonly used in DM models.
The lightest particle among $\frac{W^{a}_{0\mu} - W^{a}_{2\mu}}{\sqrt{2}}$ and $\frac{\Phi_{1} - \Phi_2}{\sqrt{2}}$ 
is stable and is a dark matter candidate for this model.

The Lagrangian of the scalar and the electroweak gauge sectors are described as
\begin{align}
& -\frac{1}{4}\sum_{j=0}^2 W^a_{j\mu\nu} W^{a \mu\nu}_j - \frac{1}{4} B_{\mu \nu} B^{\mu \nu}
+ \sum_{j=1}^2 \frac{1}{2}\tr(D^\mu \Phi_j^\dagger D_\mu \Phi_j) - V_\text{scalar}
,
\end{align}
where
\begin{align}
\label{treeV}
 V_{\text{scalar}}
=&
 m^2 H^\dagger H 
+ m_\Phi^2 \tr\left(\Phi_1^\dagger \Phi_1\right)
+ m_\Phi^2 \tr\left(\Phi_2^\dagger \Phi_2\right)+ \lambda (H^\dagger H)^2
\nonumber\\
&
+ \lambda_\Phi \left(\tr\left(\Phi_1^\dagger \Phi_1\right)\right)^2
+ \lambda_\Phi \left(\tr\left(\Phi_2^\dagger \Phi_2\right)\right)^2+ \lambda_{h\Phi} H^\dagger H  \tr\left(\Phi_1^\dagger \Phi_1\right)
\nonumber\\
&
+ \lambda_{h\Phi} H^\dagger H  \tr\left(\Phi_2^\dagger \Phi_2\right)+ \lambda_{12} \tr\left(\Phi_1^\dagger \Phi_1\right) \tr\left(\Phi_2^\dagger \Phi_2\right).
\end{align}
Some couplings are equal owing to the exchange symmetry described in Eq.~\eqref{eq:exchange-sym}. 

We assume that the scalar fields develop the following vacuum expectation values (VEVs),
\begin{align}
 \expval{H}=& \mqty(0 \\ \frac{v}{\sqrt{2}}),
\quad
 \expval{\Phi_1} = \expval{\Phi_2} = \frac{1}{\sqrt{2}}\mqty( v_\Phi & 0 \\ 0 & v_\Phi).
\label{eq:VEVs}
\end{align}
These VEVs do not break the exchange symmetry and maintain the $Z_2$ symmetry, which stabilizes the DM candidate. 
We parametrized the component fields of each scalar field as 
\begin{align}
 H=& \begin{pmatrix} i\pi_3^+ \\ \frac{
 v+ \sigma_3 - i \pi_3^0}{\sqrt{2}} \end{pmatrix},
\quad
 \Phi_j= 
   \begin{pmatrix} 
    \frac{
    v_\Phi+ \sigma_j + i \pi_j^0}{\sqrt{2}} &  i \pi_j^+ \\
    i \pi_j^-   &  \frac{
    v_\Phi + \sigma_j - i \pi_j^0}{\sqrt{2}}   
    \end{pmatrix}.
\label{eq:component_fields}
\end{align}
where $\pi^\pm_3, \pi^0_3, \pi^\pm_j, \pi^\pm_j$, and $\pi^0_j$ are would-be Nambu-Goldstone (NG) bosons.
Based on the stationary condition,
we obtain the followings: 
 \begin{align}
 \label{stationary1}
 m^2=& -\lambda v^2 - 2\lambda_{h\Phi} v_\Phi^2,\\
 \label{stationary2}
  m_{\Phi}^2=& -\frac{\lambda_{h\Phi}}{2}v^2 - (\lambda_{12} +2\lambda_\Phi)v_\Phi^2. 
 \end{align}

\subsection{Scalar boson masses}
The mass terms for scalar fields other than the would-be NG bosons are described as 
\begin{align}
- \frac{1}{2}
\mqty(\sigma_3 & \sigma_1 & \sigma_2) 
 {\cal M}_{\sigma}^2 
 \mqty(\sigma_3 \\ \sigma_1 \\ \sigma_2) 
,
\end{align}
where
\begin{align}
 {\cal M}_{\sigma}^2 
 = \mqty(
 2 \lambda v^2 & 2 v v_\Phi \lambda_{h\Phi} & 2 v v_\Phi \lambda_{h\Phi}\\
 2 v v_\Phi \lambda_{h\Phi} & 8 v_\Phi^2 \lambda_\Phi & 4 v_\Phi^2 \lambda_{12}\\
 2 v v_\Phi \lambda_{h\Phi} & 4 v_\Phi^2 \lambda_{12} & 8 v_\Phi^2 \lambda_\Phi
 ).
 \label{eq:massbosons}
\end{align}
Diagonalizing this mass matrix, we obtain the mass eigenstates denoted as $h$, $h'$, and $h_D$.
The relation between $(\sigma_3, \sigma_1, \sigma_2)$ and $(h, h', h_D)$ is
%
%
\begin{align}
\label{mix}
 \begin{pmatrix}
 \sigma_3 \\ \sigma_1 \\ \sigma_2 
\end{pmatrix}
=
R
 \begin{pmatrix}
 h \\ h' \\ h_D
\end{pmatrix}
,\end{align}
where $R$ is determined to diagonalize the mass matrix,
\begin{align}
  R^T  {\cal M}_{\sigma}^2 R
  = 
  \mqty(m_h^2 & 0 & 0 \\ 0 & m_{h'}^2 & 0 \\ 0 & 0 & m_{h_D}^2).
\end{align}
$R$ is expressed as
\begin{align}
 R=\begin{pmatrix}
  \cos\phi_h & -\sin\phi_h & 0 \\
  \frac{1}{\sqrt{2}}\sin\phi_h  & \frac{1}{\sqrt{2}}\cos\phi_h & \frac{1}{\sqrt{2}} \\
 \frac{1}{\sqrt{2}}\sin\phi_h & \frac{1}{\sqrt{2}}\cos\phi_h     & -\frac{1}{\sqrt{2}}\\
\end{pmatrix}
.
\end{align}

The couplings in the potential are expressed in terms of the masses of the physical scalars and $\phi_h$ using Eqs.~\eqref{eq:massbosons} and ~\eqref{mix},
\begin{align}
 \lambda=& \frac{m_h^2 \cos^2\phi_h + m_{h'}^2 \sin^2\phi_h}{2 v^2}, \label{eq:lambda}\\
 \lambda_{h\Phi}=& -\frac{\sin\phi_h \cos\phi_h}{2 \sqrt{2} v v_\Phi} (m_{h'}^2 - m_h^2),\\
 \lambda_{\Phi}=& \frac{m_h^2 \sin^2\phi_h + m_{h'}^2 \cos^2\phi_h + m_{h_D}^2}{16 v_\Phi^2},\\
 \lambda_{12}=& \frac{m_h^2 \sin^2\phi_h + m_{h'}^2 \cos^2\phi_h - m_{h_D}^2}{8 v_\Phi^2}.\label{eq:lambda12}
\end{align}

\subsection{Gauge boson masses}

The mass terms of the gauge bosons are given by
\begin{align}
{\cal L}\supset \begin{pmatrix}
  W_{0\mu}^+ & W_{1\mu}^+ & W_{2\mu}^+
 \end{pmatrix}
 {\cal M}_{C}^2
 \begin{pmatrix}
  W_{0}^{-\mu} \\ W_{1}^{-\mu} \\ W_{2}^{-\mu}
 \end{pmatrix}
+
 \frac{1}{2}
 \begin{pmatrix}
  W_{0\mu}^3 & W_{1\mu}^3 & W_{2\mu}^3 & B_\mu
 \end{pmatrix}
 {\cal M}_{N}^2
 \begin{pmatrix}
  W_{0}^{3\mu} \\ W_{1}^{3\mu} \\ W_{2}^{3\mu} \\ B^\mu
 \end{pmatrix}
,
\end{align}
where  
\begin{align}
 {\cal M}_{C}^2
&= 
\frac{1}{4}
\begin{pmatrix}
 g_0^2 v_{\Phi}^2 & - g_0 g_1 v_{\Phi}^2 & 0 \\
-g_0 g_1 v_{\Phi}^2 & 2 g_1^2 v_{\Phi}^2 & -g_1 g_0 v_{\Phi}^2 \\
 0  &  -g_1 g_0 v_{\Phi}^2 & g_0^2 v_{\Phi}^2
\end{pmatrix}
,\\
{\cal M}_{N}^2 
&= 
\frac{1}{4}
\begin{pmatrix}
 g_0^2 v_{\Phi}^2 & - g_0 g_1 v_{\Phi}^2 & 0 & 0 \\
-g_0 g_1 v_{\Phi}^2 & 2 g_1^2 v_{\Phi}^2 & -g_1 g_0 v_{\Phi}^2 & -g_1 g' v^2 \\
 0  &  -g_1 g_0 v_{\Phi}^2 & g_0^2 v_{\Phi}^2& 0\\
 0  &  -g_1 g' v^2 & 0 & g'^2 v^2 
\end{pmatrix}.
\end{align}
Note that $g_0 = g_2$ based on the exchange symmetry described in Eq.~\eqref{eq:exchange-sym}.
After diagonalizing these mass matrices, 
we obtain ten mass eigenstates:
$\gamma, Z, Z', W^\pm, W'^{\pm}, V^0$, and $V^{\pm}$,
where $\gamma$ is photon, $Z$ and $W^\pm$ are the electroweak gauge bosons in the SM. 

The $Z_2$-odd combinations are described as
\begin{align}
 V^0_\mu = \frac{W^3_{0\mu} - W^3_{2\mu}}{\sqrt{2}},\\
 V^\pm_\mu = \frac{W^\pm_{0\mu} - W^\pm_{2\mu}}{\sqrt{2}},
\end{align}
where $W^\pm_{j\mu} = \frac{W^1_{j\mu} \mp i W^2_{j\mu}}{\sqrt{2}}.$
The masses of $V^0$ and $V^\pm$ at the tree level are expressed as
\begin{equation}
 m_{V^{0}}^2 = m_{V^{\pm}}^2 = \frac{g_0^2 v_\Phi^2}{4} \equiv m_V.
\end{equation}
At the loop level, the electroweak interaction gives different corrections to the masses, and $V^\pm$ is slightly heavier than $V^0$~\cite{Abe:2020mph},
\begin{equation}
 m_{V^\pm} - m_{V^0} \simeq 166~\text{MeV}.
\end{equation}
This mass splitting allows $V^\pm$ to decay into $V^0$ by emitting a charged pion, and thus $V^\pm$ cannot be regarded as DM.

The full expressions of the $Z_2$-even mass eigenstates and their eigenvalues are given in \cite{Abe:2020mph} but complicated. 
Instead of showing the complicated expressions, we show their approximated expressions with $v_\Phi \gg v$, which is required by collider searches of $W'$ and the relic abundance of $V^0$ as discussed in \cite{Abe:2020mph}.
For $v_\Phi\gg v$, the masses of the $Z_2$-even gauge bosons are given by
\begin{align}
m^2_{Z'}&\simeq m^2_{W'^{\pm}}
\simeq \frac{(g_0^2+2g_1^2) v_\Phi^2}{4},\\
m^2_{W^{\pm}}&\simeq \frac{g_0^2g_1^2 v^2}{4(g_0^2+2g_1^2)},\label{eq:W-boson-mass}\\
m^2_{Z}&\simeq \frac{ g_0^2g_1^2g'^2v^2}{4e^2(g_0^2+2g_1^2)}\label{eq:Z-boson-mass},
\end{align}
where $e$ is the electric charge, expressed as $e= \left(\frac{2}{g_0^2} + \frac{1}{g_1^2} + \frac{1}{g'^2}\right)^{-\frac{1}{2}}$. The mass eigenstates of the $Z_2$-even gauge bosons are given by
\begin{align}
Z'_{\mu} &\simeq \frac{m_V}{m_{Z'}}\frac{W^3_{0\mu}+W^3_{2\mu}}{\sqrt{2}} - \sqrt{1-\left(\frac{m_V}{m_{Z'}}\right)^2} W^{3}_{1\mu},\\
W'^{\pm}_{\mu}&\simeq \frac{m_V}{m_{Z'}}\frac{W^{\pm}_{0\mu}+W^{\pm}_{2\mu}}{\sqrt{2}} - \sqrt{1-\left(\frac{m_V}{m_{Z'}}\right)^2} W^{\pm}_{1\mu},\\
W^{\pm}_{\mu}&\simeq \sqrt{1-\left(\frac{m_V}{m_{Z'}}\right)^2}\frac{W^{\pm}_{0\mu}+W^{\pm}_{2\mu}}{\sqrt{2}} + \frac{m_V}{m_{Z'}} W^{\pm}_{1\mu},\\
Z_\mu&\simeq \frac{e g_1}{\sqrt{g_0^2+2g_1^2}g'}(W^3_{0\mu}+W^3_{2\mu}) + \frac{e g_0}{\sqrt{g_0^2+2g_1^2}g'} W^{3}_{1\mu} -  \frac{e\sqrt{g_0^2+2g_1^2}}{g_0 g_1} B_\mu,\\
A_\mu& = \frac{e}{g_0}(W^{3}_{0\mu}+W^{3}_{2\mu}) + \frac{e}{g_1} W^{3}_{1\mu} + \frac{e}{g'} B_\mu,
\end{align}
where 
\begin{align}
\label{g0g1}
g_0&\simeq\sqrt{2}g_W\frac{ m_{Z'}}{m_V}\frac{ 1}{\sqrt{\frac{m_{Z'}^2}{m_V^2}-1}},\\
\label{g1g0}
g_1&\simeq g_W \frac{ m_{Z'}}{m_V}.
\end{align}
Here, $g_W$ is defind by $g_W = \left( \frac{2}{g_0^2}+ \frac{1}{g_1^2}\right)^{-\frac{1}{2}}$, and it is found that $g_W \simeq 0.65$ for $v_\Phi \gg v$~\cite{Abe:2020mph}.
Using this $g_W$, the electric charge is expressed as $e = \qty(\frac{1}{g_W^2} + \frac{1}{g'^2})^{-\frac{1}{2}}$. 
Also, $m_{W^\pm}$ and $m_Z$ are expressed as $m_{W^\pm} = \frac{g_W v}{2}$ and $m_Z = \frac{\sqrt{g_W^2 + g'^2}}{2} v$. Therefore, $g_W$ is interpret as the SU(2)$_L$ gauge coupling.

\subsection{Some couplings of the vector dark matter}
In the model, the $V^0$ and $V^\pm$ couplings to the SM gauge bosons $(\gamma, Z, W^\pm)$ are essential for DM phenomenology. 
They allow the annihilation processes such as $V^0 V^0 \to W^+ W^-$, and the annihilation processes determine the relic abundance of DM via the freeze-out mechanism.
The triple-gauge-boson couplings including $V^{0,\pm}$ and the SM gauge bosons are given by
\begin{align}
 {\cal L}
\supset&
i g_{V^+V^-\gamma}
\left(
  (V^{-\nu} \lrpdv_\mu V^+_\nu ) A^{0\mu}
+ (V^{+\nu} \lrpdv_\mu A^0_\nu ) V^{-\mu}
+ (A^{0\nu} \lrpdv_\mu V^-_\nu ) V^{+\mu}
\right)
\nonumber\\
&
+
i g_{V^+V^-Z}
\left(
  (V^{-\nu} \lrpdv_\mu V^+_\nu ) Z^{\mu}
+ (V^{+\nu} \lrpdv_\mu Z_\nu ) V^{-\mu}
+ (Z^{\nu} \lrpdv_\mu V^-_\nu ) V^{+\mu}
\right)
\nonumber\\
&
+
i g_{W^+V^-V^0}
\left(
  (V^{-\nu} \lrpdv_\mu W^+_\nu ) V^{0\mu}
+ (W^{+\nu} \lrpdv_\mu V^0_\nu ) V^{-\mu}
+ (V^{0\nu} \lrpdv_\mu V^-_\nu ) W^{+\mu}
\right)
\nonumber\\
&
+
i g_{V^+W^-V^0}
\left(
  (W^{-\nu} \lrpdv_\mu V^+_\nu ) V^{0\mu}
+ (V^{+\nu} \lrpdv_\mu V^0_\nu ) W^{-\mu}
+ (V^{0\nu} \lrpdv_\mu W^-_\nu ) V^{+\mu}
\right)
,
\end{align}
where $A\lrpdv_\mu B = A \partial_\mu B - (\partial_\mu A) B$, and 
\begin{align}
 g_{V^+V^-\gamma} =& e,\\
                  g_{V^-V^+Z} \simeq& g_W \frac{m_W}{m_Z} \simeq g_{WWZ}^\text{SM},\\
 g_{V^-W^+V^0} = g_{W^-V^+V^0} \simeq& g_W
.
\end{align}
These couplings are proportional to the SU(2)$_L$ gauge coupling $g_W$. 
This result allows us to interpret $V^a$ as an SU(2)$_L$ triple.

\subsection{Parameters}

There are nine parameters in the electroweak sector,
\begin{align}
\left( g_0, g_1, g', m^2, m_\Phi^2, \lambda, \lambda_\Phi, \lambda_{h\Phi}, \lambda_{12} \right).
\end{align}
Instead of using them as input parameters, we adopt the following set:
\begin{align}
\left(e, v, m_{h}, m_Z, m_V, m_{Z'}, m_{h'}, m_{h_D}, \phi_h \right),  
\end{align}
where 
$v$ is relate to the Fermi constant $G_F$ as $v= (\sqrt{2} G_F)^{-1/2}$.
The values of the first four parameters are already known.
The remaining five parameters $(m_V, m_{Z'}, m_{h'}, m_{h_D}, \phi_h)$ are treated as the model parameters.  
In the rest of this subsection, we discuss constrains on them.

The lightest $Z_2$-odd particle is the DM candidate for this model. 
Because $m_{V^\pm} > m_{V^0}$ by the loop effect, the lighter one of $V^0$ and $h_D$ is the DM candidate in this model. 
Following~\cite{Abe:2020mph}, we assume that $m_{V^0} < m_{h_D}$ and treat $V^0$ as the DM candidate.

As discussed in~\cite{Abe:2020mph},
this model can explain the measured value of the DM energy density through the freeze-out mechanism for 3~TeV $\lesssim m_{V^0} \lesssim 19$~TeV,
which is larger than $m_W$ and $m_Z$. 
This heavy $V^0$ is realized by the large hierarchy between the two VEVs. Therefore, we investigate the model with $v_\Phi \gg v$.

\subsection{Constraints on $\phi_h$}

Higgs couplings to the SM particles give a constraint on $\phi_h$. The $h$ coupling to $W$ boson $g_{WWh}$, and to the SM fermions $g_{ffh}$, are given by 
\begin{align}
    g_{WWh} =& g_{WWh}^\text{SM} \cos\phi_h,\\ 
    g_{ffh} =& g_{ffh}^\text{SM} \cos\phi_h,
\end{align}
where $g_{WWh}^\text{SM}$ and
$g_{ffh}^\text{SM}$ are the couplings in the SM given by $g_{WWh}^\text{SM} = \frac{2m_W^2}{v}$ and 
$g_{ffh}^\text{SM} = \frac{m_f}{v}$. 
The ATLAS and CMS experiments measure those couplings~\cite{ATLAS:2022vkf, CMS:2022dwd}.
The authors of~\cite{Abe:2020mph} used the ATLAS results~\cite{ATLAS:2019nkf} and concluded that $\abs{\phi_h}< 0.3$.
Recent results from the ATLAS experiment are as follows~\cite{ATLAS:2022vkf}: 
\begin{align}
    \frac{g_{WWh}}{g_{WWh}^\text{SM}} =&1.035 \pm 0.031,\\
    \frac{g_{ffh}}{g_{ffh}^\text{SM}} =&0.95 \pm 0.05,
\end{align}
with a positive correlation of 39\%. 
Using this constraint on the couplings, 
we obtain $\abs{\phi_h} < 0.24$.
Similarly, 
using the results from the CMS experiment~\cite{CMS:2022dwd}, we find $\abs{\phi_h} < 0.23$.

DM direct detection experiments provide bounds on $\phi_h$.
The DM candidate $V^0$ scatters off the nuclei by exchanging scalar bosons in $t$-channel. 
The scattering cross section is proportional to $\sin^2 2\phi_h$.
The current upper bound on the cross section requires $\abs{\phi_h} \lesssim \mathcal{O}(0.1)$. 
The XENONnT experiment~\cite{XENON:2015gkh} can probe the region of parameter space for $\abs{\phi_h} \gtrsim \mathcal{O}(0.01)$ in the future~\cite{Abe:2020mph}.  

Another constraint on $\phi_h$ comes from the perturbative unitarity bounds on the scalar quartic couplings.
It gives the similar upper bound on $\phi_h$ as the current DM direct detection experiment, $\abs{\phi_h} \lesssim 0.1$~\cite{Abe:2020mph}.
Hence, small $\abs{\phi_h}$ is required from both the experimental and theoretical bounds.

We focus on $\abs{\phi_h} < \mathcal{O}(0.01)$, where both the theoretical and the experimental bounds are absent. 
The direct detection experiments cannot probe the model with such a small $\abs{\phi_h}$ value in the near future. 
Hence, other experiments, such as GW detections, are useful for examining the model. 
To be specific, we take $\phi_h=0.001$ in the following analysis,
which was chosen as a benchmark point in \cite{Abe:2020mph}.
Note that the relic abundance is unchanged even if we take much smaller $\phi_h$ because $V^{0}$ and $V^{\pm}$ mainly annihilate into the SM gauge bosons and these processes are independent of $\phi_h$. Therefore, in the following analysis of the scalar potential, we take $\phi_h = 0$ as an approximation of the small $\phi_h$ value.

\subsection{Constraints on $m_{Z'}$}
Searches for heavy vector bosons in the ATLAS and CMS experiments give the constraints on $m_{Z'}$.  
In this model, $W'$ couples to the SM fermions, and its coupling is given by
\begin{align}
    g_{W'ff} \simeq -\sqrt{\frac{m_{Z'}^2}{m_V^2}-1} g_{Wff}^\text{SM},
\end{align}
where $g_{Wff}^\text{SM}$ is the $W$ coupling to the SM fermions. 
Owing to this coupling, $W'$ can be searched by the $pp\to W'\to \ell\nu$ process. 
The ATLAS and CMS experiments provide bounds on $m_{W'}$ with $g_{W'ff} = g_{Wff}^\text{SM}$~\cite{ATLAS:2019lsy,CMS:2018hff}.
In this model, $g_{W'ff} \neq g_{Wff}^\text{SM}$, and thus the authors in~\cite{Abe:2020mph} recasted the bound and found that $m_{W'}\gtrsim$ 7~TeV if $m_{V^0} \lesssim 4$~TeV~\cite{Abe:2020mph}.
They also discussed the prospect at the HL-LHC~\cite{ATLAS:2018tvr} and found that 
the $W'$ search can test the model for $m_{Z'} \lesssim 8.5$~TeV and $m_{V^0} \lesssim 5.8$~TeV. 

%
%
%
%

\section{Loop and finite temperature corrections to scalar potential and phase transition}
\label{sec:potential}

In this section, we discuss the phase transitions of the scalar potential. We calculate the finite-temperature effective potential and obtain the conditions necessary for the first-order phase transition. 

There are two distinct energy scales: the electroweak scale denoted as $v$, and the scale characterizing the dark sector, represented as $v_\Phi$.
As discussed in Section~\ref{sec:model}, 
the two scales are well separated, $v_\Phi \gg v$. 
Consequently, the phase transition within the dark sector precedes the electroweak phase transition. 
During this transition within the dark sector, electroweak symmetry is restored. 
This clear separation of the two scales allows us to discuss the two phase transitions separately.

The phase transition in the dark sector can be first-order because the typical values of $g_0$ and $g_1$ are $\mathcal{O}(1)$ in a large region of the parameter space~\cite{Abe:2020mph} and provide sizable corrections to the scalar potential. 
In the following, we focus on phase transition in the dark sector.

Before moving forward, we briefly discuss the phase transition in the electroweak sector.
It is expected to be the same as that in the SM, which is not first-order. Because the two scales are well separated, $v_\Phi \gg v$, 
the electroweak phase transition can be discussed using an effective theory in which all the new fields are integrated out, and the Lagrangian is provided by the SM together with higher dimensional operators.
In this case, a first-order electroweak phase transition occurs if the energy scale of the new physics is $\mathcal{O}(100)$~GeV~\cite{Grojean:2004xa,Hashino:2022ghd}. However, the energy scale of the new physics in this model is much higher than the electroweak scale, $v_\Phi \sim \mathcal{O}(1)$~TeV. Therefore, the electroweak phase transition in this model is not first-order; thus, we do not further investigate it in this paper.

\subsection{Effective potential}\label{app:potential}

\begin{figure*}[t]
\centering
\includegraphics[width=0.6\textwidth]{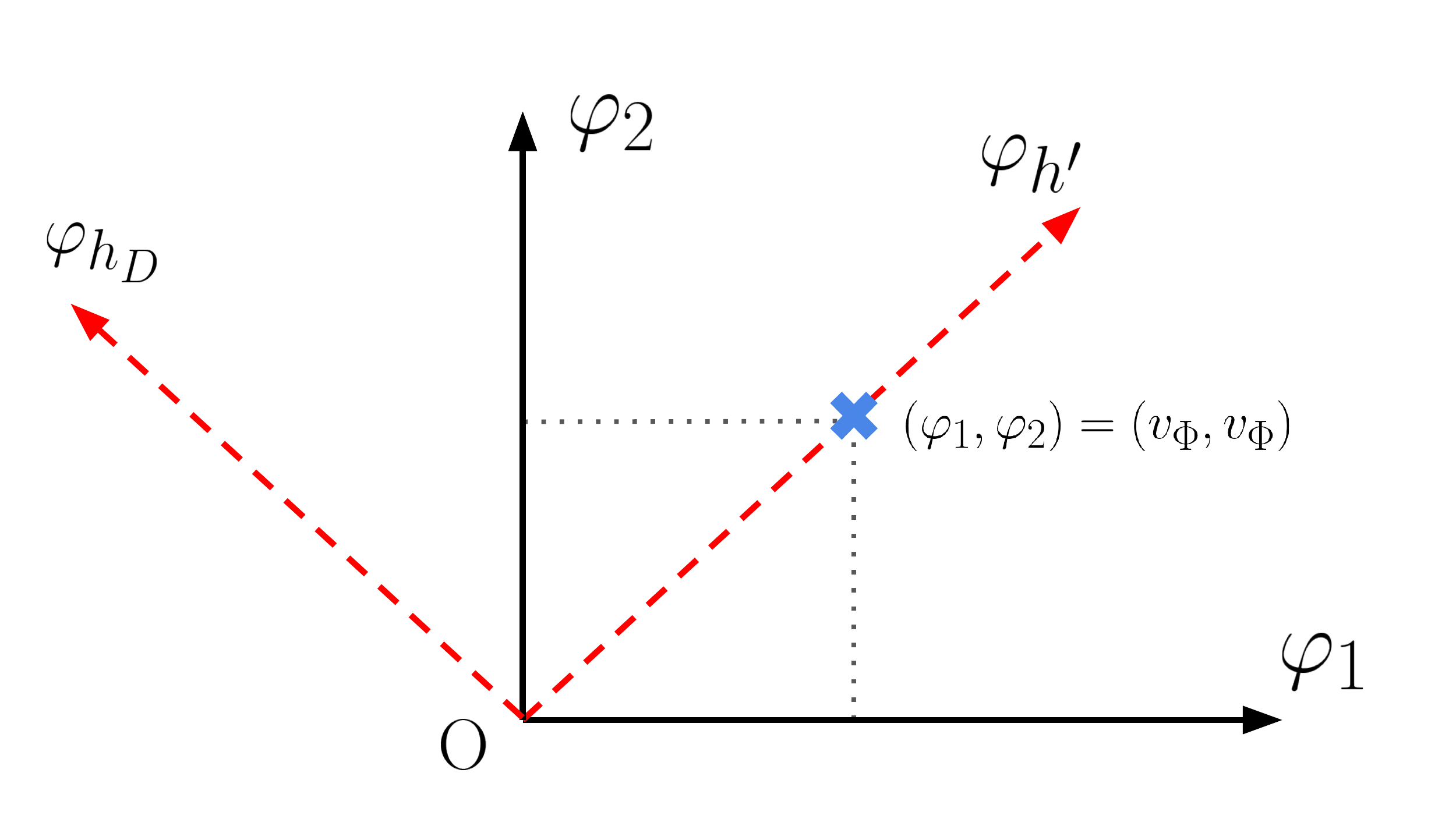}
\caption{
The solid black axes represent the values of the classical fields $\Phi_{1,2}$, whereas the red dashed axes represent $h'$ and $h_D$.
The true vacuum, which corresponds to $\varphi_{1}=\varphi_{2}=v_\Phi$, is indicated by the blue cross mark.
\label{fig:axis}
}
\end{figure*}

During the phase transition in the dark sector, 
the electroweak symmetry is well restored, as discussed at the beginning of this section, 
and the classical background fields of the scalar fields are expressed as 
\begin{align}
\label{cl}
 \expval{\Phi_i}_{cl}=  \frac{1}{\sqrt{2}}\begin{pmatrix} \varphi_i & 0 \\ 0 & \varphi_i \end{pmatrix},
 \quad
 \expval{H}_{cl} = \mqty(0 \\ 0),
\end{align}
where $i=1,2$. 
We define $\varphi_{h'}$ and $\varphi_{h_D}$ as 
\begin{align}
\label{classical2}
 \varphi_{h'} = \frac{1}{\sqrt{2}}\left(\varphi_{1}+\varphi_{2}\right),
 \quad  
 \varphi_{h_D} = \frac{1}{\sqrt{2}}\left(\varphi_{1}-\varphi_{2}\right).
\end{align}
They correspond to the classical fields of $h'$ and $h_D$ described in Eq.~\eqref{mix} and shown in Fig.~\ref{fig:axis}.

At the one-loop level, the effective potential including thermal corrections is expressed as
\begin{align}
\label{VeffwithT}
 V_{\text{eff}}(\varphi_{h'},\varphi_{h_D}, T)
=& 
m_\Phi^2 (\varphi_{h'}^2+\varphi_{h_D}^2) + \frac{\lambda_\Phi}{2} (\varphi_{h'}^4+6\varphi_{h'}^2\varphi_{h_D}^2+\varphi_{h_D}^4) + \frac{\lambda_{12}}{4}(\varphi_{h'}^2-\varphi_{h_D}^2)^2
\nonumber\\
& + V_\text{CW}(\varphi_{h'},\varphi_{h_D}) 
+ \Delta V_T(\varphi_{h'},\varphi_{h_D}, T) 
+ \delta V (\varphi_{h'},\varphi_{h_D}), 
\end{align}
where $V_\text{CW}$ represents the Coleman-Weinberg potential at the one-loop level, 
$\Delta V_T(\varphi_{h'},\varphi_{h_D}, T)$ denotes the thermal correction, and $\delta V$ contains counterterms that cancel UV divergence.

To evaluate $V_\text{eff}$, we must expand the fields around the classical background fields and obtain the field-dependent masses, denoted as $\overline{m}_j^2$. 
The mass terms of the fluctuation fields are expressed as follows\footnote{
As elucidated in Section~\ref{sec:model}, we focus on $\abs{\phi_h} < \mathcal{O}(0.01)$ and set $\phi_h \simeq 0$. 
At this limit, $\lambda_{h\Phi}$ vanishes and does not appear here. 
As discussed later in this paper, essential contributions are from the gauge couplings, and thus, the results are not significantly modified even if $\lambda_{h\Phi}$ is considered in the effective potential. 
}: 
\begin{align}
    \mathcal{L}
    \supset&
    -\frac{1}{2} \mqty(\bar{h}_1 & \bar{h}_2) \overline{\mathcal{M}}_s^2 \mqty(\bar{h}_1  \\ \bar{h}_2) 
    -\frac{1}{2} \sum_{i=1}^2 \sum_{a=1}^3 \overline{m}_{\pi_i}^2 (\bar{\pi}_i^a)^2 
    +\frac{1}{2} \sum_{a=1}^{3}\mqty(\bar{W}_0^a &\bar{W}_1^a & \bar{W}_2^a) 
     \overline{{\cal M}}_G^2 
     \mqty(\bar{W}_0^a \\ \bar{W}_1^a \\ \bar{W}_2^a),
\end{align}
where 
\begin{align}
    \overline{m}_{\pi_1}^2 =& 2m_\Phi^2 + 4\lambda_{\Phi} \varphi_1^2 + 2\lambda_{12} \varphi_2^2,\\
    \overline{m}_{\pi_2}^2 =& 2m_\Phi^2 + 2\lambda_{12} \varphi_1^2   + 4\lambda_{\Phi} \varphi_2^2,\\
    \overline{\mathcal{M}}_s^2 
    =& \mqty(
       2 m_\Phi^2 + 12 \lambda_\Phi \varphi_1^2 + 2 \lambda_{12}\varphi_2^2 
    &  4 \lambda_{12} \varphi_1 \varphi_2
    \\ 4 \lambda_{12} \varphi_1 \varphi_2
    &  2 m_\Phi^2 + 12 \lambda_\Phi \varphi_2^2 + 2 \lambda_{12}\varphi_1^2 
    ),\\
    \overline{{\cal M}}_G^2
        =& 
        \frac{1}{4}
        \mqty(
         g_0^2 \varphi_1^2 & - g_0 g_1 \varphi_1^2 & 0 \\
        -g_0 g_1 \varphi_1^2 & g_1^2 (\varphi_1^2 + \varphi_2^2) & -g_1 g_0 \varphi_2^2 \\
         0  &  -g_1 g_0 \varphi_2^2 & g_0^2 \varphi_2^2
        )
    .
\end{align}
We obtain the eigenvalues of $\overline{\mathcal{M}}_s^2$ as follows: 
\begin{align}
    \overline{m}_{h'}=&
    2 m_\Phi^2 + (\lambda_{12} + 6\lambda_\Phi)(\varphi_1^2 + \varphi_2^2)
    - \sqrt{
      (\lambda_{12} + 6\lambda_\Phi)^2 (\varphi_1^2 - \varphi_2^2)^2 
    + 16 \lambda_{12}^2 \varphi_1^2 \varphi_2^2
    },\\
    \overline{m}_{h_D}^2
    =&
    2 m_\Phi^2 + (\lambda_{12} + 6\lambda_\Phi)(\varphi_1^2 + \varphi_2^2)
    + \sqrt{
      (\lambda_{12} + 6\lambda_\Phi)^2 (\varphi_1^2 - \varphi_2^2)^2 
    + 16 \lambda_{12}^2 \varphi_1^2 \varphi_2^2
    },
\end{align}
and the eigenvalues of $\overline{\mathcal{M}}_G^2$ as follows: 
\begin{align}
    \overline{m}_{W}^2 =& 0,\\
    \overline{m}_{V}^2
    =& \frac{g_0^2+g_1^2}{8}(\varphi_1^2 + \varphi_2^2)
    - \frac{1}{8} \sqrt{ (g_0^2+g_1^2)^2(\varphi_1^2 - \varphi_2^2)^2 + 4 g_1^4 \varphi_1^2 \varphi_2^2 },\\
    \overline{m}_{Z'}^2
    =& \frac{g_0^2+g_1^2}{8}(\varphi_1^2 + \varphi_2^2)
    + \frac{1}{8} \sqrt{ (g_0^2+g_1^2)^2(\varphi_1^2 - \varphi_2^2)^2 + 4 g_1^4 \varphi_1^2 \varphi_2^2 }.
\end{align}
Using these field-dependent masses, we obtain the following: 
\begin{align}
\label{divCW}
V_\text{CW}
=& 
\frac{1}{64\pi^2} \sum_j n_j \overline{m}_j^4
\left(\log\frac{\overline{m}_j^2}{\mu^2} -c_j-\frac{1}{\bar{\epsilon}}\right),
\end{align}
where $1/\bar{\epsilon}$ represents UV divergence, 
$c_j = \frac{3}{2} \left(\frac{5}{6}\right)$ for the scalar (gauge) bosons, 
$\mu$ is a renormalization scale, 
the summation is taken for the particles coupling to $\Phi_1$ or $\Phi_2$ ($j=\pi_1, \pi_2, h', h_D, W, V$, and $Z'$), and $n_j$ represents the degrees of freedom of the particles: 
\begin{align}
  & n_{\pi_1} = n_{\pi_2} = 3, \nonumber \\
  & n_{h'}= n_{h_D}= 1, \nonumber \\
  & n_{W}=n_{V}=n_{Z'} = 9.
\end{align}
We use $\mu = m_V$ in the following analysis.
%
%
%
%
%
%
The counterterms are described as follows: 
\begin{align}
\label{counter}
 \delta V
=&
   \delta \Lambda
 + \delta m_\Phi^2 (\varphi_{h'}^2+\varphi_{h_D}^2) 
 + \frac{\delta\lambda_\Phi}{2} (\varphi_{h'}^4 + 6\varphi_{h'}^2\varphi_{h_D}^2+\varphi_{h_D}^4) 
 + \frac{\delta\lambda_{12}}{4} (\varphi_{h'}^2-\varphi_{h_D}^2)^2,
\end{align}
where $\delta \Lambda$, $\delta m_\Phi^2$, $\delta\lambda_\Phi$, and $\delta \Lambda$ are determined using the following renormalization conditions \cite{Anderson:1991zb,Delaunay:2007wb}:\footnote{Hereafter, we focus only on the real part of the effective potential.} 
	\begin{align} 
		 \label{renormalization1} 
   \left.V_\text{eff.}\right|_{\varphi_{h'} = \varphi_{h_D} = 0} 
   &= 0,\\ 
     \label{renormalization2}
		  \left.\frac{\partial V_\text{eff} }{\partial\varphi_{h'}} \right|_{\varphi_{h'}=\sqrt{2}v_\Phi, \varphi_{{h_{D}}}=0}     
    &= 0,\\ 
		\label{renormalization3} 
  \left.\frac{\partial^2 V_\text{eff} }{\partial\varphi_{h'}^2} \right|_{\varphi_{h'}=\sqrt{2}v_\Phi, \varphi_{{h_{D}}}=0}
  &= m_{h'}^2 - \Delta\Sigma_{h'h'},\\
		\label{renormalization4}  
   \left.\frac{\partial^2  V_\text{eff}}{\partial\varphi_{h_{D}}^2} \right|_{\varphi_{h'}=\sqrt{2}v_\Phi, \varphi_{h_{D}}=0}
  &= m_{h_D}^2 - \Delta\Sigma_{h_{D}h_{D}},
	\end{align}
where 
$m_{h'}$ and $m_{h_D}$ are the pole masses for $h'$ and $h_D$, respectively, $\Delta\Sigma_{ii}=\Sigma_{ii}(p^2=m_{i}^2)-\Sigma_{ii}(p^2=0)$, and $\Sigma_{ii}(p^2)$ is the two-point function of $h'$ and $h_D$.
These are evaluated using \texttt{FeynArts}, \texttt{FormCalc}, and \texttt{LoopTools}~\cite{Hahn:2000kx, Hahn:1998yk}.
The field-independent divergent terms in $V_\text{eff}$ are absorbed using Eq.~\eqref{renormalization1}.
The result is slightly lengthy and is presented in Appendix~\ref{app:Veff-at-T=0}.

The thermal loop correction $\Delta V_T$ is expressed as follows~\cite{Dolan:1973qd}: 
	\begin{align} 
	\label{finiteT}
	 \Delta V_T&= \frac{T^4}{2\pi^2} 
  \Biggl\{ \sum_{j}
  n_j  \int_0^\infty d x x^2\ln 
  \left[ 1- \exp \left( -\sqrt{x^2+a_j^2}\right) \right] \Biggl\},
	\end{align}
where $a_j^2 = \dfrac{\overline{m}_j^2}{T^2}$.
It is noteworthy that the SM fermions do not contribute to the effective potential of the dark sector.

The thermal loop calculations result in IR divergences. 
To avoid this, we replace the field-dependent masses in the effective potential with the following~\cite{Parwani:1991gq,Arnold:1992rz}: 
\begin{align}
    \label{eq:w2mass0}
    \overline{m}_{\pi_{1,2}}^2 
    \to& 
    \overline{m}_{\pi_{1,2}}^2 
    + T^2 \left( 2\lambda_{\Phi} + \frac{2\lambda_{12}}{3}+ \frac{3(g_0^2+g_1^2)}{16}\right)
    ,\\ 
    \label{eq:w2mass1}
    \overline{m}_{h', h_D}^2 
    \to& 
    \overline{m}_{h', h_D}^2 
    +  T^2 \left( 2\lambda_{\Phi} + \frac{2\lambda_{12}}{3}+ \frac{3(g_0^2+g_1^2)}{16}\right)
    ,\\
    \overline{\mathcal{M}}_G^2
    \to& 
    \overline{\mathcal{M}}_G^2
    + a^{\rm (L)} 
    \mqty( \frac{5T^2g_0^2}{6} & 0 & 0 \\
    0 & \frac{13T^2g_1^2}{6} & 0 \\
    0 & 0 & \frac{5T^2g_0^2}{6})
    ,\label{eq:w2mass}
\end{align}
where $a^{(\text{L})} = 1$ for the longitudinal mode and $a^{(\text{L})} = 0$ for the transverse mode, respectively.

\subsection{First-order phase transition}\label{subsec:FOFT}

We investigate the temperature-dependence of the effective potential.
%
%
\begin{figure*}[t]
\centering
\includegraphics[width=0.47\textwidth]{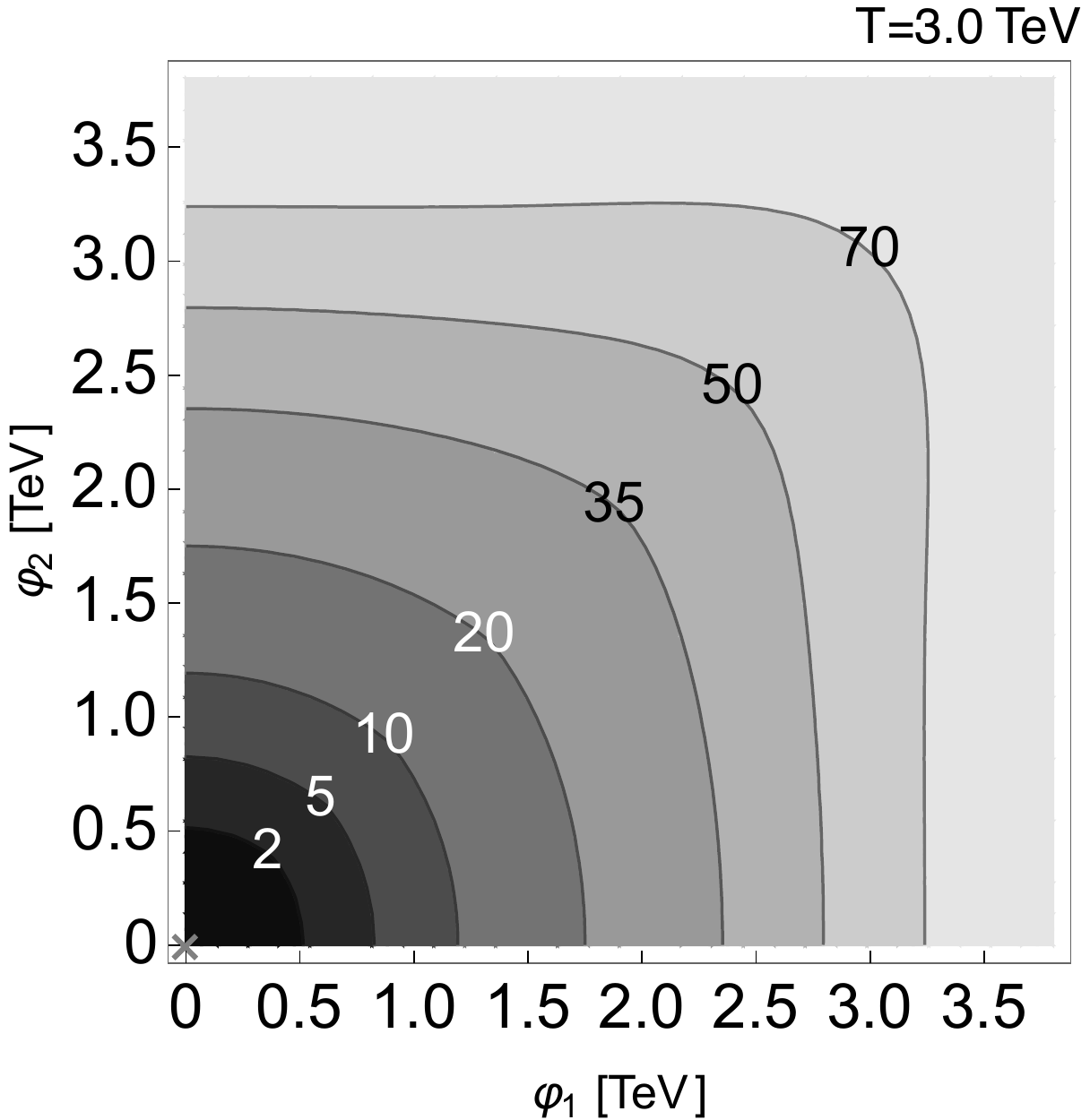}
\includegraphics[width=0.47\textwidth]{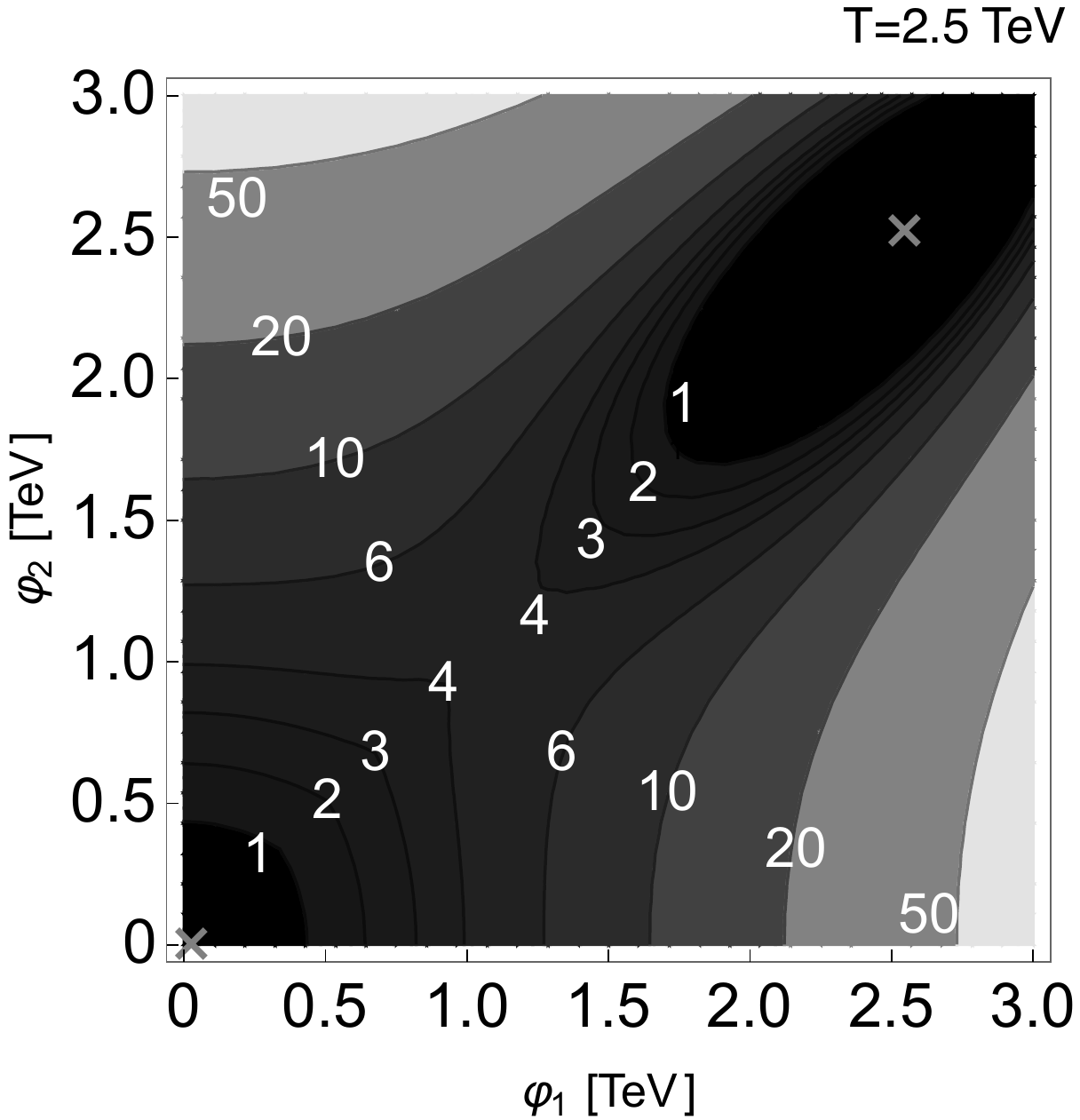}
\caption{
Contours of $(V_{\rm eff}(\varphi_1,\varphi_2,T)-V_{\rm eff}(0,0,T))/V_{\rm eff}(0,0,T)$ with $m_{Z'}=5.0$~TeV, $m_V=3$~TeV, $m_{h'}=0.7 m_V$, and $m_{h_D}=1.2 m_V$ at $T=3.0$ and 2.5~TeV. 
The horizontal and vertical axes are the $\varphi_1$ and $\varphi_2$ axes, respectively. 
Darker regions correspond to lower potential values.
The cross marks denote the locations of the global minima.
At $T=2.5$~TeV, two degenerate minima emerge, one at the origin and the other at $\varphi_1=\varphi_2 \simeq 2.55$~TeV.
\label{fig:Tevolutions}
}
\end{figure*}
Figure~\ref{fig:Tevolutions} shows the contours of $(V_{\rm eff}(\varphi_1,\varphi_2,T)-V_{\rm eff}(0,0,T))/V_{\rm eff}(0,0,T)$ at two different temperatures: $T=3.0$ and 2.5~TeV. 
Here, we choose $m_{Z'}=5.0$~TeV, $m_V=3$~TeV, $m_{h'}=0.7 m_V$, and $m_{h_D}=1.2 m_V$ for the sake of illustration.
In this figure, the darker regions correspond to lower potential values.
At $T=3.0$~TeV, the potential has only one minimum at $\varphi_1=\varphi_2=0$, and thus the symmetry is restored. 
At $T=2.5$~TeV, there are two minima. One is at $\varphi_1=\varphi_2=0$, and the other is at $\varphi_1 = \varphi_2 \simeq 2.55$~TeV.
The two minima have the same potential energies but are separated. 
Below this temperature, the potential has only one global minimum at $\varphi_1=\varphi_2 \neq 0$.
For a given value of $\sqrt{\varphi_1^2 + \varphi_2^2}$, the potential takes minimum value at $\varphi_1 = \varphi_2$.
As a result, the first-order phase transition occurs in the potential along the line $\varphi_1 = \varphi_2$.
The critical temperature ($T_C$) is the temperature at which the potential has two separate minima that have the same potential energy. In the example here, $T_C =2.5$~TeV.
We define $\varphi_C$ by the value of $\sqrt{\varphi_1^2 + \varphi_2^2}$ at the location of the second minimum, namely the minimum not at the origin, of the potential at $T_C$. In Fig.~\ref{fig:Tevolutions}, we find $\varphi_C \simeq 3.61$~TeV and $\varphi_C/T_C \simeq 1.44$, which is sufficient to produce detectable GW spectra as we discuss in Section~\ref{sec:numerical}. 

%
%

\begin{figure*}[t]
\centering
\includegraphics[width=0.45\textwidth]{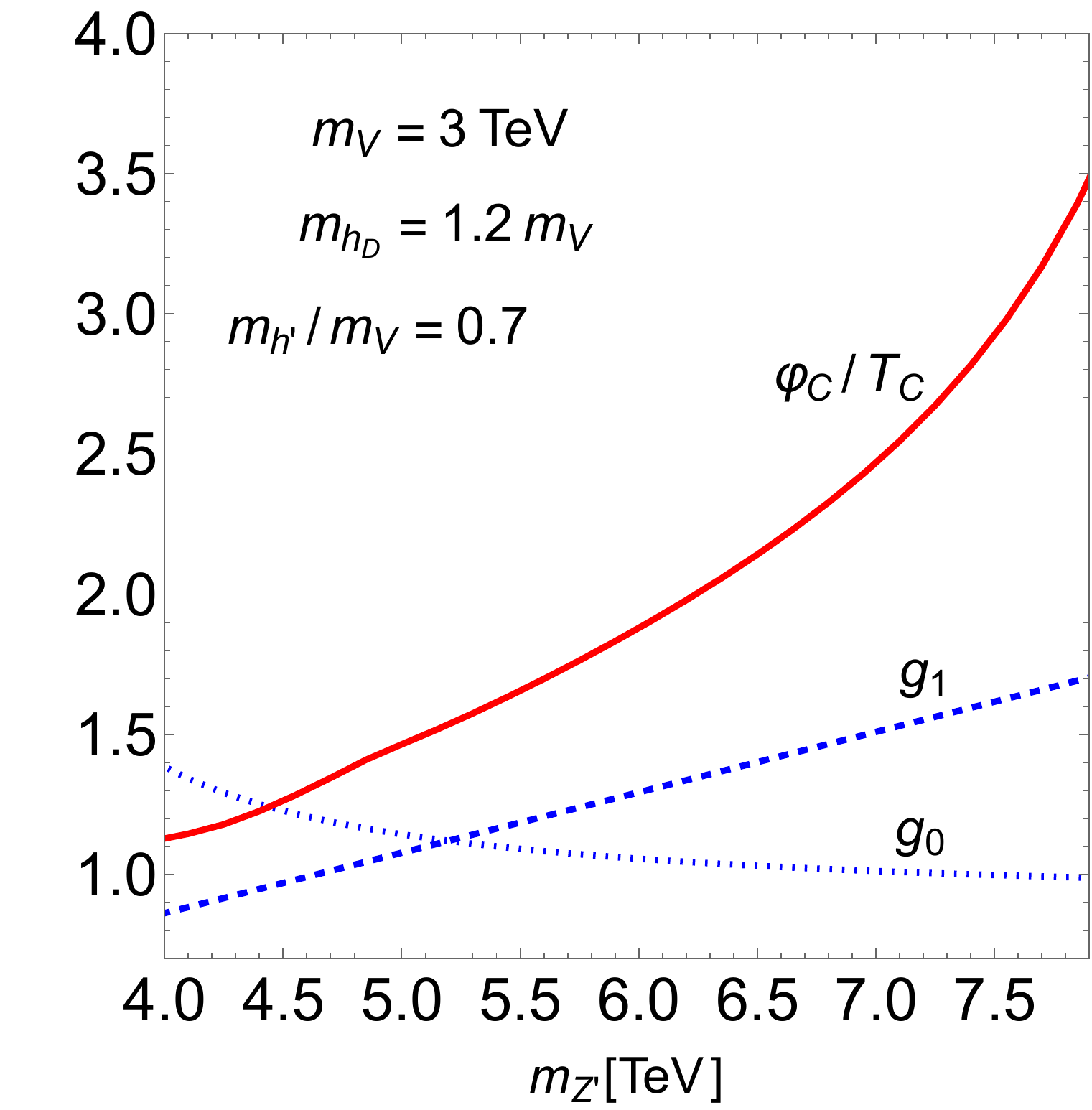}
\caption{
Values of $g_0$, $g_1$, and $\varphi_C/T_C$ for $m_V=3$~TeV, $m_{h_D}=1.2 m_V$, and $m_{h'}/m_V=0.7$. 
The blue dotted curve, blue dashed lines, and red solid curve indicate the values of $g_0$, $g_1$, and $\varphi_C/T_C$, respectively. 
\label{fig:g0g1VCTC}
}
\end{figure*}
 We discuss the dependence of $\varphi_C/T_C$ on $m_{Z'}$. 
 We calculate $V_\text{eff}(\varphi_1, \varphi_2, T)$ for $m_V=3$~TeV, $m_{h_D}=1.2 m_V$, $m_{h'}/m_V=0.7$, with varying the value of $m_{Z'}$ from 3~TeV to 9~TeV. We find that the phase transition of $V_\text{eff}$ is the first-order with these parameters, and thus $\varphi_C$ and $T_C$ are obtained.
 Figure~\ref{fig:g0g1VCTC} shows the values of $\varphi_C/T_C$ as a function of $m_{Z'}$. 
 The values of the gauge couplings are also shown in the figure, which are evaluated from Eqs.~\eqref{g0g1} and ~\eqref{g1g0}. 
 It is shown that an increase in the gauge couplings $g_0$ or $g_1$ leads to an increase in the $\varphi_C/T_C$ value. 
 This is because the thermal loop effects of bosons intensify as the gauge coupling increases.
 In the regions characterized by a large value of $\varphi_C/T_C$, the observable GW spectra may be produced by the first-order phase transition~\cite{Ahriche:2018rao}.  
The values of these gauge couplings are relatively large for obtaining the measured value of the DM energy density~\cite{Abe:2020mph}. Therefore, the model realizes a strong first-order phase transition over a broad range of parameter space.

\begin{figure*}[t]
\centering
\includegraphics[width=0.45\textwidth]{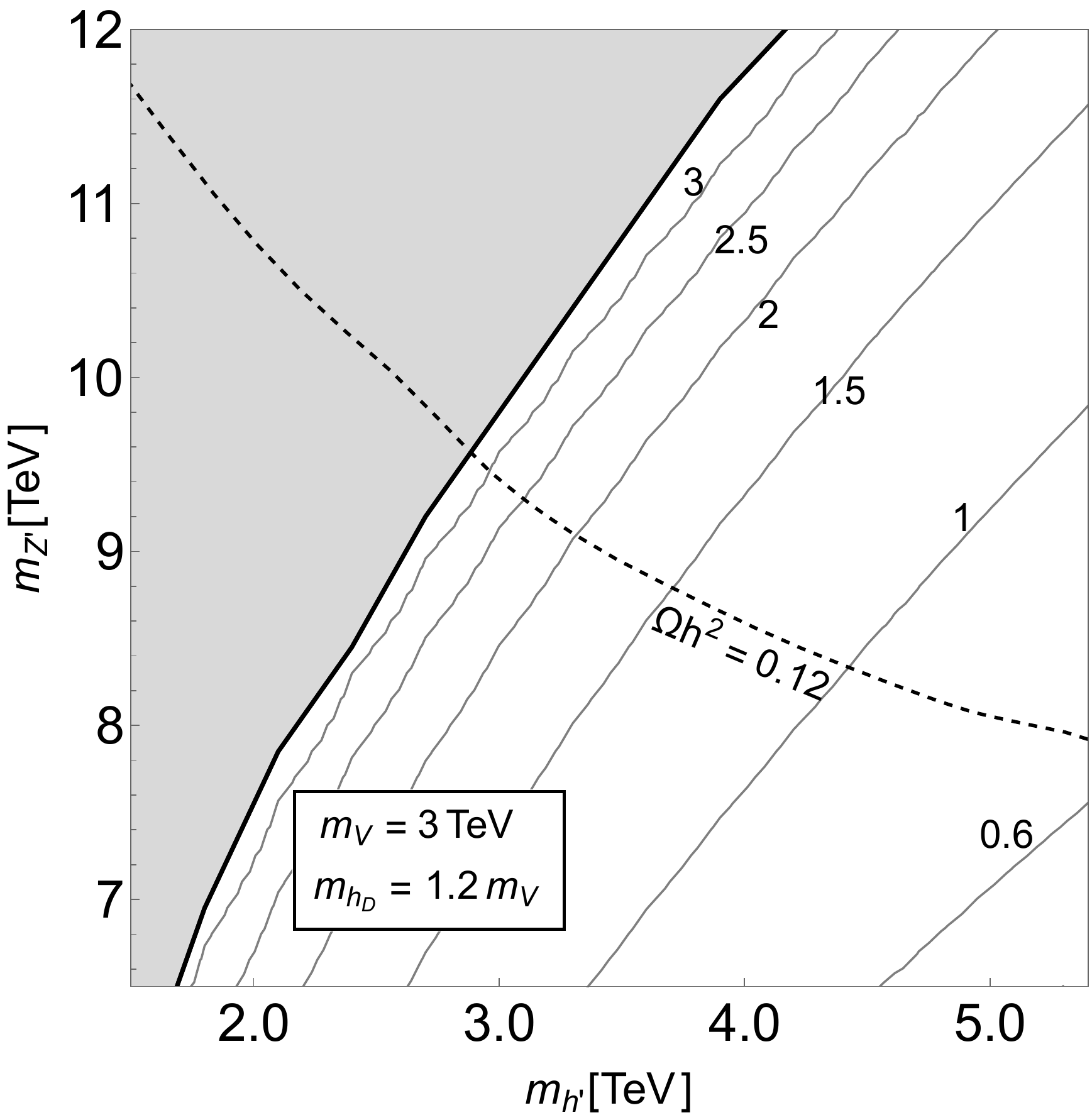}~~~~
\includegraphics[width=0.45\textwidth]{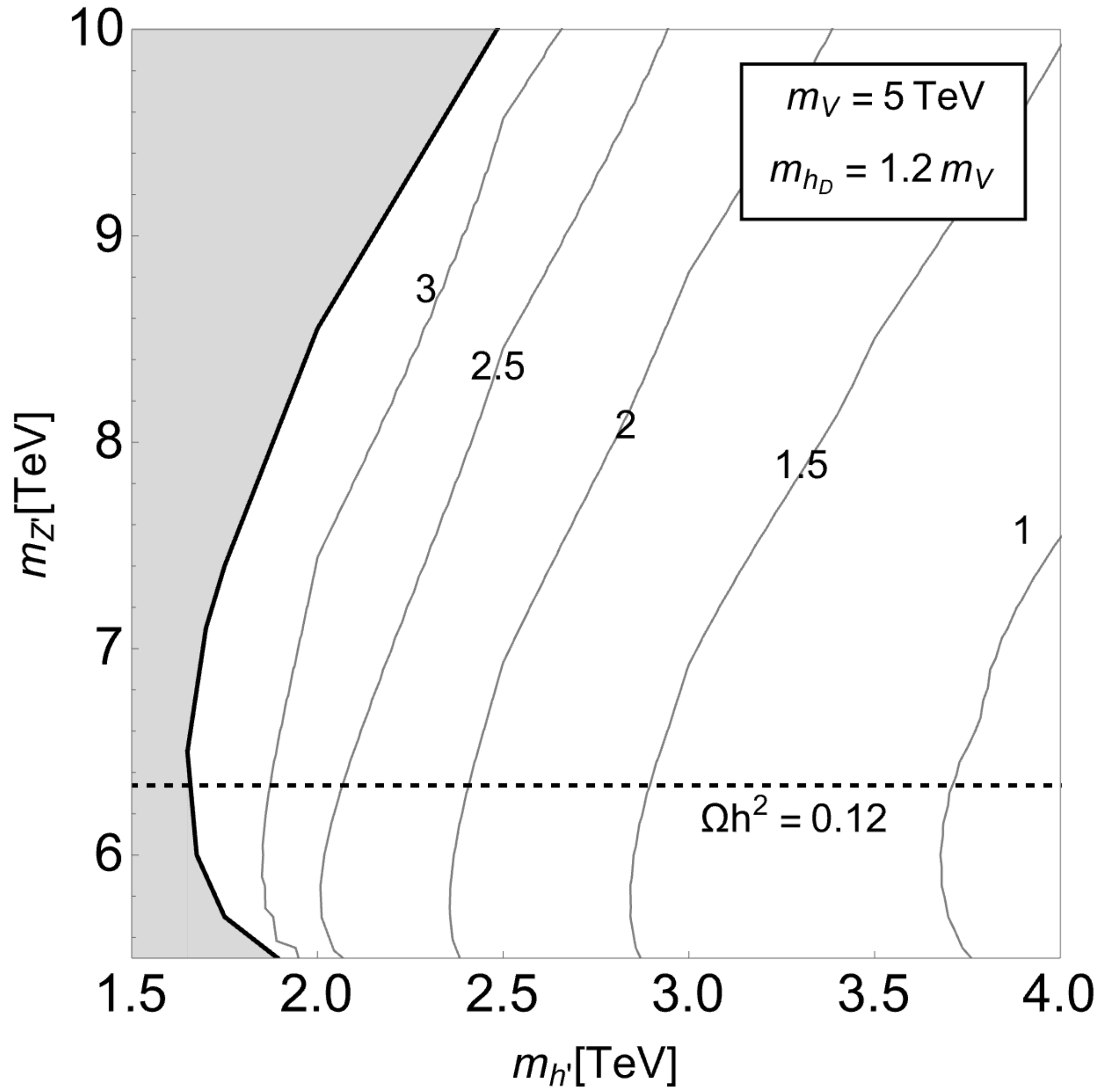}
\includegraphics[width=0.45\textwidth]{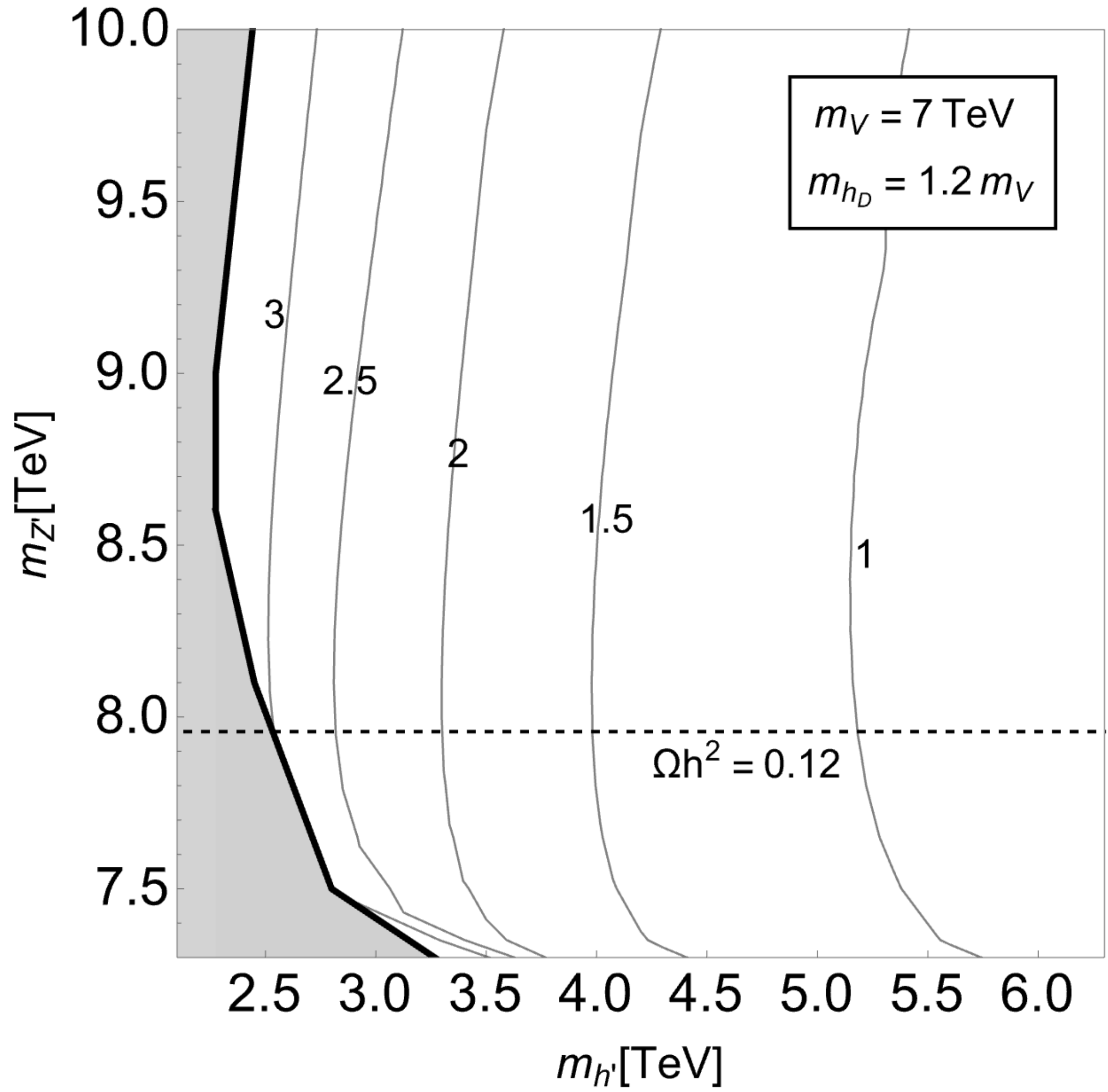}
\caption{
Values of $\varphi_C/T_C$ in the $m_{h'}$-$m_{Z'}$ plane with $m_V=3$, 5 and 7~TeV and $m_{h_D}=1.2 m_V$. 
The upper left (right) panel shows the value of $\varphi_C/T_C$ for $m_V=3$ (5)~TeV. 
The lower panel shows the value for $m_V=7$~TeV. 
The black dashed lines represent $\Omega h^2=0.12$.
 The vacuum expectation value remains zero in the gray regions, even in the current universe. 
\label{fig:VCTC}
}
\end{figure*}
The value of $\varphi_C/T_C$ also depends on $m_{h'}$. 
Figure~\ref{fig:VCTC} shows the values of $\varphi_C/T_C$ in the $m_{h'}$-$m_{Z'}$ plane for $m_V=3$, 5, and 7~TeV.  We set $m_{h_D}=1.2 m_V$ in all three panels. 
The black dashed lines represent $\Omega h^2=0.12$.  
In the gray region, 
the vacuum expectation value remains zero, even in the current universe. 
The values of $\varphi_C/T_C$ are larger for smaller $m_{h'}$ regions. It is understood from the shape of the effective potential around the origin as follows. With the renormalization conditions given in Eqs.~\eqref{renormalization1}--\eqref{renormalization4}, the effective potential at $T = 0$ is expressed as 
\begin{align}
 V_\text{CW}+\delta V
 \simeq - \frac{1}{4} 
 \left(
 m_{h'}^2
 - \frac{9(m_V^4+m_{Z'}^4)}{16\pi^2 v_\Phi^2} 
 \right)
 (\varphi_1^2 + \varphi_2^2)
,
\label{eq:Veff_T=0_around-origin}
\end{align}
for small $\varphi_1^2 + \varphi_2^2$.
The coefficient is negative at the tree level but 
can be positive at the loop level because of gauge boson contributions. If it is positive, a local minimum, not a local maximum, is generated at $\varphi_1 = \varphi_2 = 0$ even at the tree level.  
The renormalization condition given in 
Eq.~\eqref{renormalization2} guarantees that there is another local minimum at $\varphi_1 = \varphi_2 \neq 0$. 
As a result, two local minimum emerge 
if the coefficient in Eq.~\eqref{eq:Veff_T=0_around-origin} is positive,  
and a potential barrier is generated at the tree level; thus, a first-order phase transition occurs at a finite but small temperature. This is why $\varphi_C/T_C$ values are larger for smaller $m_{h'}$ values. We also find that
the energy difference between the two local minima can be approximately described as
\begin{align}
 \eval{V_\text{eff}}_{\varphi_1 = \varphi_2 = v_\Phi, T=0} 
 -  \eval{V_\text{eff}}_{\varphi_1 = \varphi_2 = 0, T=0} 
\simeq&
 - \frac{v_\Phi^2}{4} 
\left(
 m_{h'}^2
 - \frac{9(m_V^4 + m_{Z'}^4)}{32\pi^2 v_\Phi^2}
\right)
.
\label{eq:V-V0}
\end{align}
If this value is positive, the vacuum at $\varphi_1 = \varphi_2 = 0$ is the global minimum at $T = 0$. 
Subsequently, a first-order phase transition does not occur for small $m_{h'}$; hence, 
 the gray regions appear in Fig.~\ref{fig:VCTC}. 
Even if Eq.~\eqref{eq:V-V0} is negative, namely the vacuum at $\varphi_1  = \varphi_2  \neq 0$ is the global minimum, the field values may keep $\varphi_1 = \varphi_2 = 0$ if a tunneling rate is low, as we discuss in Section~\ref{sec:GW}. 
This is another reason for 
the gray regions.

\begin{figure*}[tbp]
\centering
\includegraphics[width=0.32\textwidth]{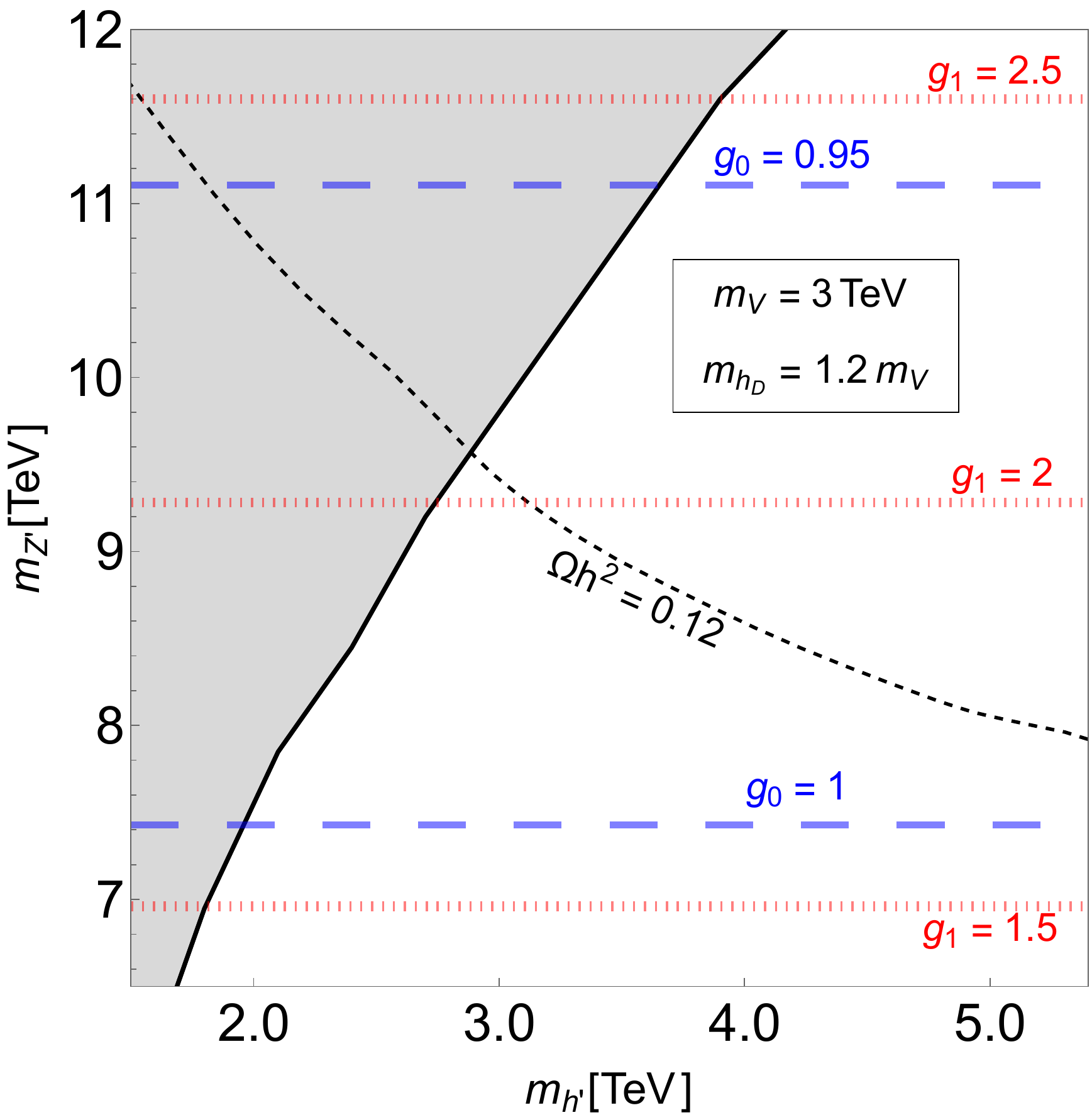}
\includegraphics[width=0.32\textwidth]{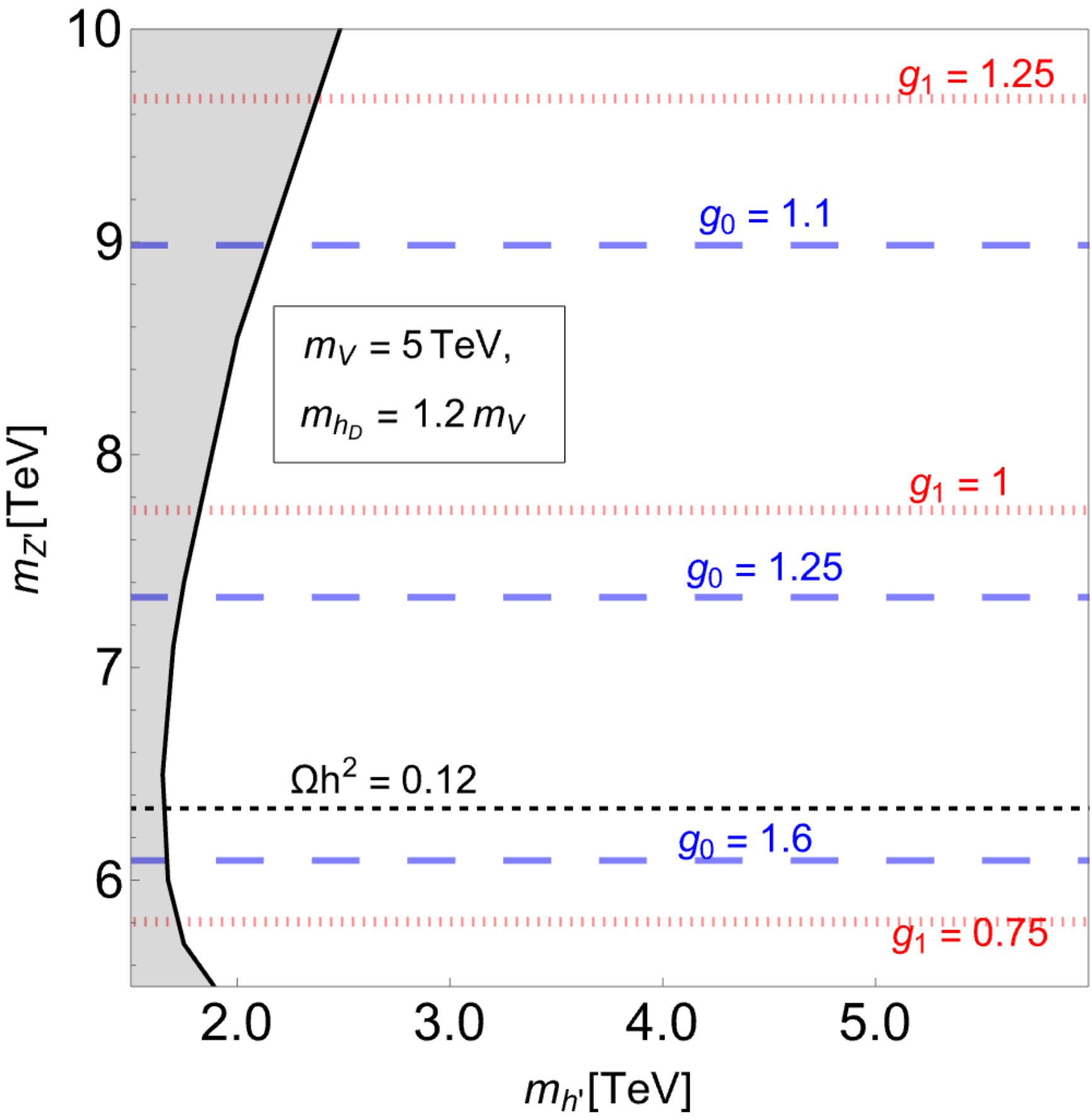}
\includegraphics[width=0.33\textwidth]{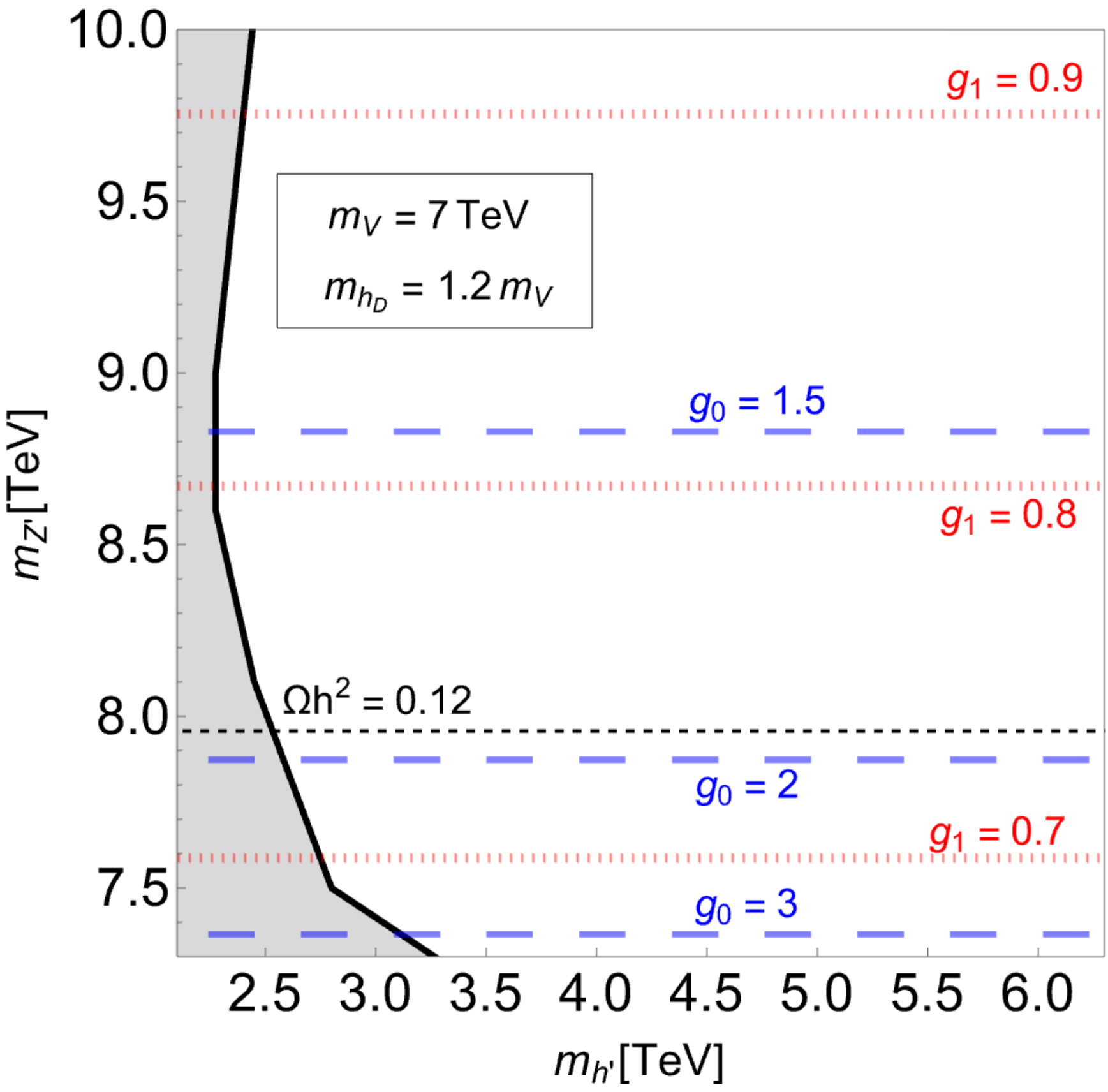}
 \caption{
 The blue-dashed lines represent the values of $g_0$, 
 and the red-dotted lines represent the values of $g_1$. 
 The parameter choice and other color notations are the same as in Fig.~\ref{fig:VCTC}.
}
\label{fig:g0_g1}
\end{figure*}

We calculate the values of the gauge couplings in Figs.~\ref{fig:VCTC}. 
The values of $g_0$ and $g_1$ with $m_V=3$, 5 and 7~TeV are shown in Fig.~\ref{fig:g0_g1}. 
We find that they are at most $~2.5$, and thus the loop factor $g_{0,1}^2/(4\pi)^2$ is at most $2.5^2/(4\pi)^2 \sim 0.04$. Therefore, our perturbative calculation is valid.

\section{Gravitational wave spectrum}
\label{sec:GW}

We use the GW observational experiments to explore the model parameter space in which a strong first-order phase transition occurs. 
Before discussing the testability of the model through GW, 
we briefly introduce herein the evolution of the GW spectrum from the first-order phase transition, which is characterized by (i) nucleation temperature $T_n$,
(ii) $\alpha$, which is a parameter related to the latent heat,
(iii) $\beta$, the inverse of the duration of the phase transition,
and (iv) a bubble wall velocity $v_b$.

The nucleation temperature $T_n$ is determined using the following: 
	\begin{equation} 
 \label{eq:GamH}
 \eval{\frac{\Gamma}{H^4}}_{T=T_n}=1,
	\end{equation}
where $H$ is the Hubble parameter, and $\Gamma$ is the critical bubble nucleation rate per unit volume per unit time defined as~\cite{Linde:1981zj}
\begin{equation}
\label{eq:decay}
\Gamma (T) \simeq  T^4 \left(\frac{S_3}{2\pi T}\right)^{\frac{3}{2}} \exp \left(-S_3/T\right), 
 \end{equation}
where $S_3$ is a three-dimensional Euclidean action~\cite{Coleman:1977py}. 
It is expressed as 
\begin{equation}
  S_3 = \int d^3r \left[\frac{1}{2}  (\vec{\nabla}\varphi_b)^2
  + V_{\rm eff}^{}(\varphi_b, T)\right],
\end{equation}
where $\varphi_b$ is the bounce solution. 
It is obtained by solving the following differential equation: 
\begin{equation}  
\frac{d^2 \varphi^{}}{d r^2}+\frac{2}{r} \frac{d \varphi^{}}{d r} - \frac{\partial V_{\rm eff}^{}}{\partial \varphi^{}} =0 ,
\end{equation}
with the following boundary conditions: 
\begin{equation}
\left. \frac{d\varphi^{}}{dr}\right|_{r=0} = 0, \quad \lim_{r\to \infty} \varphi = 0.
\end{equation}
If $\Gamma/H^4<1$ today, the phase transition would not be completed even in the current universe. 
This inconsistency necessitates that $\Gamma/H^4$ must have reached one before today.
We calculate the Hubble parameter using $H=8\pi^3g_* T^4/90m_{\rm Pl}^2$, where $m_{\rm Pl}$ is the Planck mass, and $g_*$ is the degrees of freedom in the plasma. 

\begin{figure*}[t]
\centering
\includegraphics[width=0.47\textwidth]{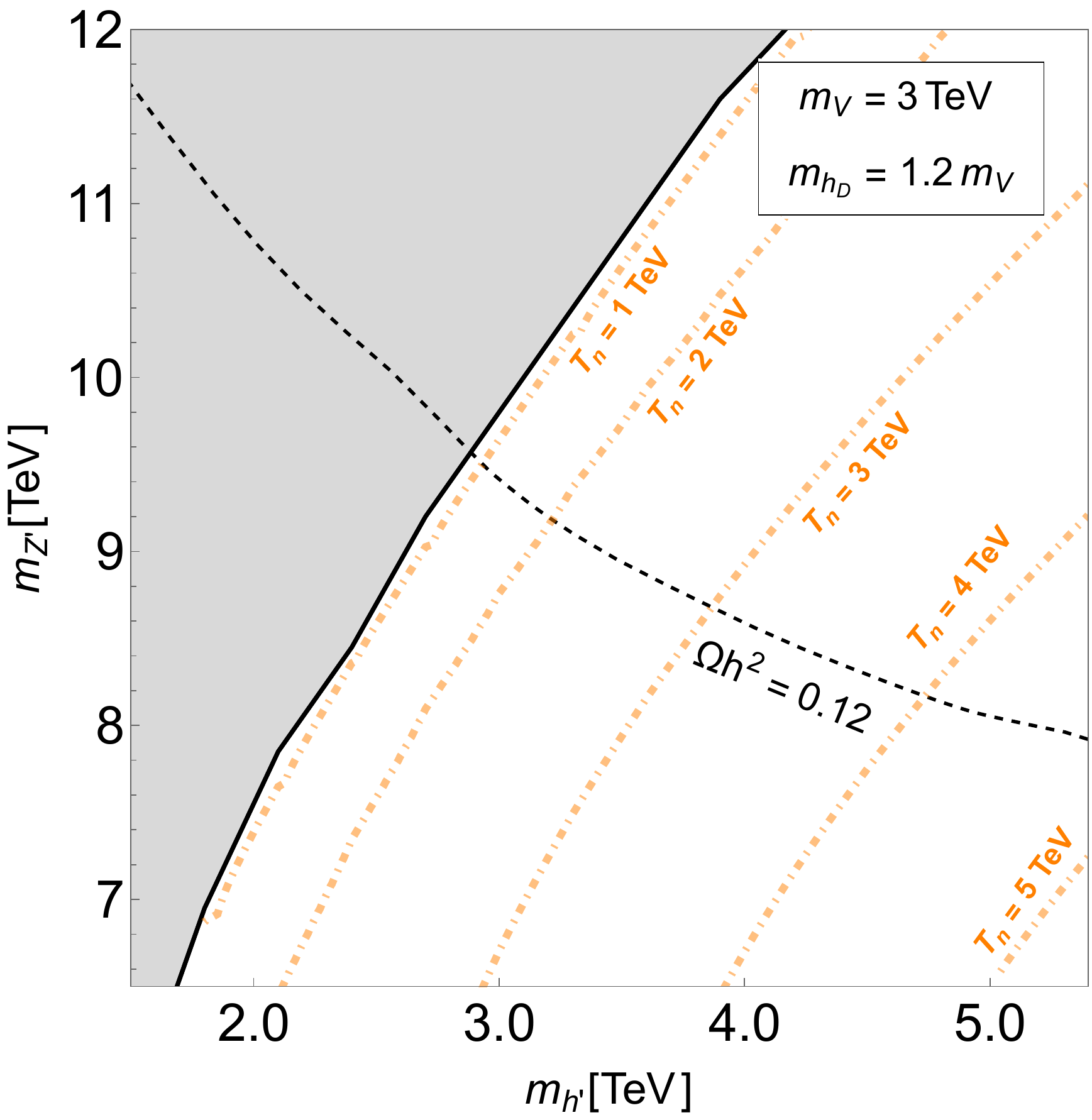}
\includegraphics[width=0.47\textwidth]{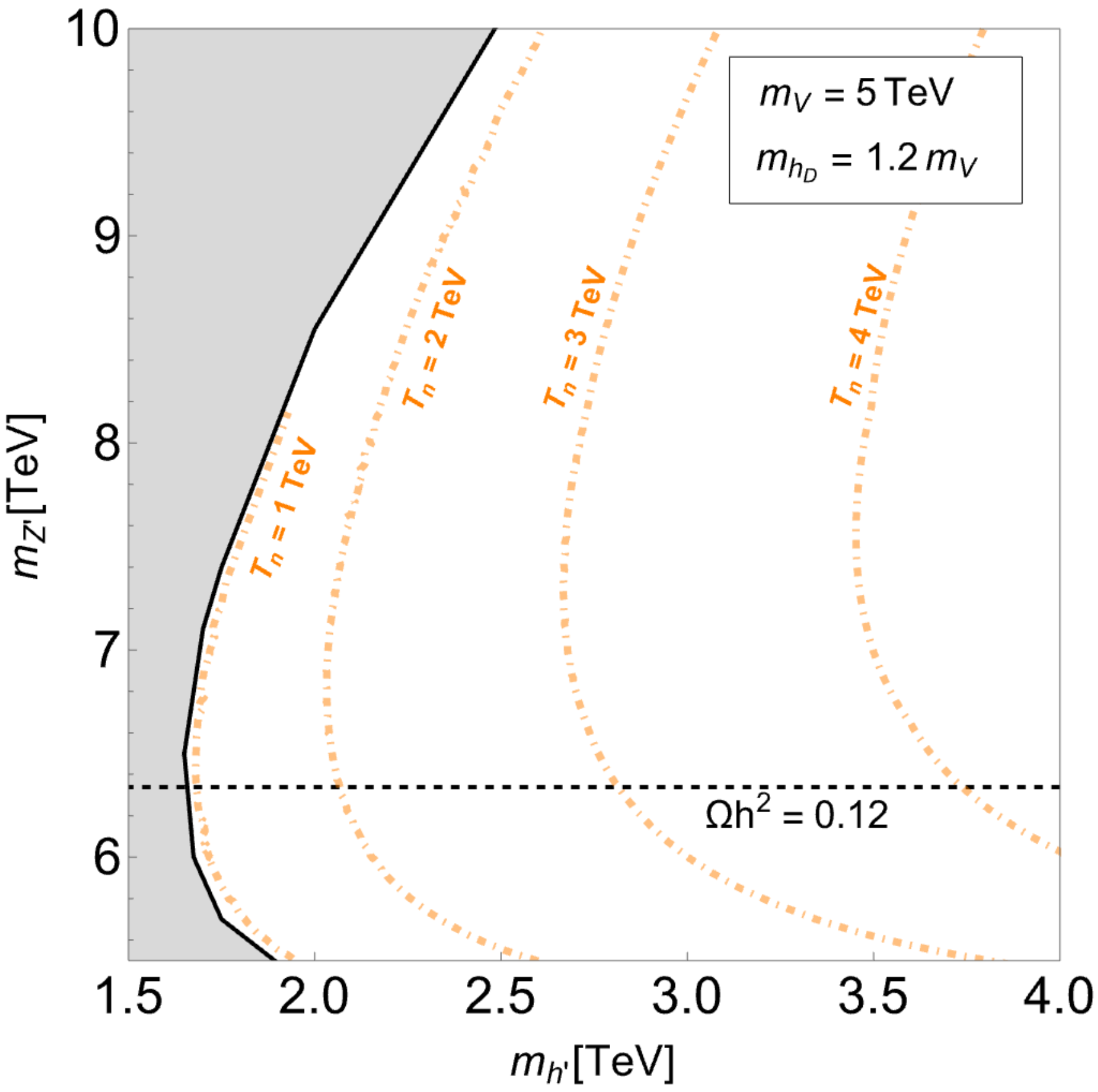}
\includegraphics[width=0.47\textwidth]{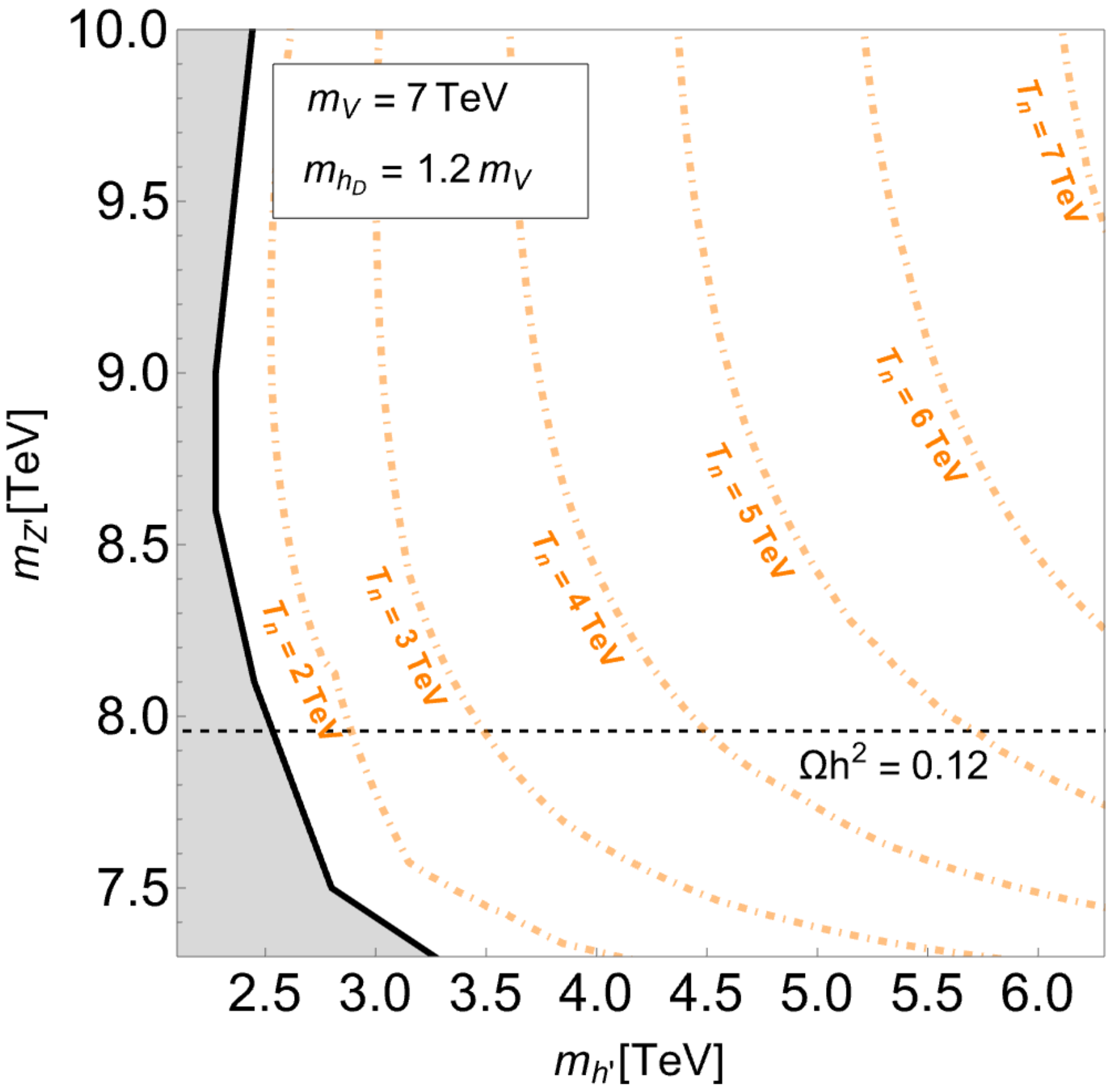}
\caption{
Values of $T_n$ in the $m_{h'}$-$m_{Z'}$ plane with $m_V=3$, 5 and 7~TeV and $m_{h_D}=1.2 m_V$. 
The orange dashed-dotted lines are the values of $T_n$. 
The upper left (right) panel shows the value of $T_n$ for $m_V=3$ (5)~TeV. 
The lower panel shows the value for $m_V=7$~TeV. 
Other color notations are the same as in Fig.~\ref{fig:VCTC}.
}
\label{fig:Tn}
\end{figure*}
As discussed at the beginning of Section~\ref{sec:potential}, the phase transition within the dark sector precedes the electroweak phase transition. 
This indicates that the value of $T_n$ of the phase transition temperature within the dark sector is higher than that of the electroweak transition, which is about 100 GeV. 
Figure~\ref{fig:Tn} shows the value of $T_n$ of the phase transition in the dark sector. 
%
We find find $T_n \gtrsim 1$~TeV. So, the phase transition in the dark sector is well separated from the electroweak transition.

The second parameter, denoted as $\alpha$, represents the released latent heat normalized by the radiative energy density, $\rho_{\rm rad}(T)=(\pi^2/30)g_* T^4$ at the nucleation temperature $T_n$. It is defined as
	\begin{align} 
	\label{definitiona}
	\alpha = \epsilon(T_n)/ \rho_{\rm rad}(T_n),
	\end{align}
where $\epsilon(T)$ represents the released latent heat and is defined as 
     \begin{align}
     \label{latenth}
      \epsilon(T)
      =  \Delta V_{\rm eff} -T
      \frac{\partial  \Delta V_{\rm eff} }{\partial T},
    \end{align}
with
    \begin{align}
     \Delta V_{\rm eff} =  V_{\rm eff}(\varphi_-(T),T) - V_{\rm eff}(\varphi_+(T),T).
    \end{align}
Here, $\varphi_{+}$ and $\varphi_{-}$ are the order parameters for the broken and unbroken phases, respectively.

The third parameter, denoted as $\beta/H$, represents the inverse of the duration of the phase transition.
It is expressed as 
	\begin{align} 
		\label{definitionb}
	\frac{\beta}{H}\equiv T_n\left.\frac{d}{dT}\left(\frac{S_3}{T}\right)\right|_{T=T_n}.
	\end{align}

The fourth parameter, $v_b$, is the bubble-wall velocity in the rest frame of the plasma far from the wall. 
We choose $v_b=0.3$ and $1$ as benchmark points in our numerical analysis.\footnote{
There is no easy way to estimate bubble wall velocity; however, the authors of~\cite{LiLi:2023dlc} have recently proposed a formula to estimate terminal wall velocity for strongly coupled first-order phase transition. 
Therefore, it is generally better to consider velocity as a free parameter, and we use $v_b=0.3$ and $1$ as benchmark points in our analysis.}

The first-order phase transition generates GWs through three mechanisms: bubble collision, plasma turbulence, and compression waves in the plasma.
The contribution of bubble wall collisions becomes significant when the bubble walls accelerate rapidly~\cite{Bodeker:2009qy}. 
However, runaway bubble wall expansion is unlikely~\cite{Bodeker:2017cim}.  
The contribution of the turbulence is described in~\cite{Kosowsky:2001xp,Hindmarsh:2015qta}.
However, it is generally subdominant. 
In the following analysis, we consider the GW from the compression wave. 
The fitting function for the numerical simulations of the GW spectrum is expressed as~\cite{Caprini:2019egz,Hindmarsh:2017gnf} 
%
	\begin{align}
  \Omega_{\rm GW} (f) 
  &= 2.061 F_{\rm gw,0}\tilde{\Omega}_{\rm gw} \left(\frac{f}{\tilde{f}_{\rm comp}}\right)^3
  \left(\frac{7}{4+3(f/\tilde{f}_{\rm comp})^2}\right)^{7/2}\notag
  \\
  &\times\begin{cases}
  \left(\frac{\kappa_v\alpha}{1+\alpha}\right)^2  (H_*R_*)
   &\left(H_*R_* \leq \sqrt{\frac{3}{4}\kappa_v\alpha/(1+\alpha)}\right)
   \\
    \left(\frac{\kappa_v\alpha}{1+\alpha}\right)^{3/2} (H_*R_*)^2 &\left(\sqrt{\frac{3}{4}\kappa_v\alpha/(1+\alpha)} < H_*R_*\right)
  \end{cases},  \label{eq:GWspcom} 
	  \end{align}
where $F_{\rm gw,0} = 3.57 \times 10^{-5} \left(100/g_\ast\right)^{1/3}$, $\tilde{\Omega}_{\rm gw}=1.2 \times 10^{-2}$, $H_*R_*=(8\pi)^{1/3}(\beta/H)^{-1} $ max($c_s,v_b$), and $\tilde{f}_{\rm comp}$ is the peak frequency expressed as
	\begin{align}
  \tilde{f}_{\rm comp} \simeq 0.26 \left(H_*R_*\right)^{-1}
  \left(\frac{T_n}{{\rm GeV}}\right)
  \left(\frac{g_\ast}{100}\right)^{1/6} ~{\rm Hz}.
	\end{align}
In addition, $\kappa_v$ in the fitting function is an efficiency factor~\cite{Espinosa:2010hh}:
\begin{align}
  \kappa_v(v_b, \alpha)\simeq
  \left\{
  \begin{array}{ll}
    \frac{ c_s^{11/5}\kappa_A \kappa_B }{(c_s^{11/5} -  v_b^{11/5} )\kappa_B
      +  v_b c_s^{6/5} \kappa_A} & \left( v_b \lesssim c_s \right) \\
     \kappa_B + ( v_b - c_s) \delta\kappa 
    + \frac{( v_b - c_s)^3}{ (v_J - c_s)^3} [ \kappa_C - \kappa_B -(v_J - c_s) 
    \delta\kappa ]
    \quad &\left( c_s <  v_b < v_J \right)\\
     \frac{ (v_J - 1)^3 v_J^{5/2}  v_b^{-5/2}
      \kappa_C \kappa_D }
    {[( v_J -1)^3 - ( v_b-1)^3] v_J^{5/2} \kappa_C
      + ( v_b - 1)^3 \kappa_D } &
     \left( v_J \lesssim v_b \right)
  \end{array}\right.,    
\end{align}
where $c_s=0.577$, and 
\begin{align}
  \kappa_A \simeq& v_b^{6/5} \frac{6.9 \alpha}{1.36 - 0.037 \sqrt{\alpha} + \alpha},\\
  \kappa_B \simeq& \frac{\alpha^{2/5}}{0.017+ (0.997 + \alpha)^{2/5} },\\
  \kappa_C \simeq& \frac{\sqrt{\alpha}}{0.135 + \sqrt{0.98 + \alpha}},\\
  \kappa_D \simeq& \frac{\alpha}{0.73 + 0.083 \sqrt{\alpha} + \alpha},\\
  v_J=& \frac{\sqrt{2/3\alpha +\alpha^2}+\sqrt{1/3}}{1+\alpha},\\
  \delta \kappa\simeq& -0.9 \ln \frac{\sqrt{\alpha}}{1+\sqrt{\alpha}}.
\end{align}
Using these equations, the GW spectrum of the first-order phase transition is evaluated.
To discuss the detail of the testability of the model using the GW spectrum, we use the signal-to-noise ratio (SNR)~\cite{Seto:2005qy}, which is expressed as 
\begin{align}
\label{eq:SNR}
\mathrm{SNR}=\sqrt{\mathcal{T}\int_{f_\mathrm{min}}^{f_\mathrm{max}}\mathrm{d}f\left[\frac{h^2\Omega_\mathrm{GW}(f)}{h^2\Omega_\mathrm{sen}(f)}\right]^2},
\end{align}
where $\mathcal{T}$ is the experimental period, and $h^2\Omega_\mathrm{sen}(f)$ is a sensitivity curve of a GW detector. 
Three future GW experiments are considered: LISA, DECIGO, and BBO~\cite{Yagi:2011wg,Klein:2015hvg}.  
A GW spectrum with SNR $ >10$ can be used to explore the model parameters~\cite{Caprini:2015zlo,Hashino:2022ghd}. 
We use this criterion to discuss the testability of the GW observations in our numerical analysis.

\begin{figure*}[t]
\centering
\includegraphics[width=0.41\textwidth]{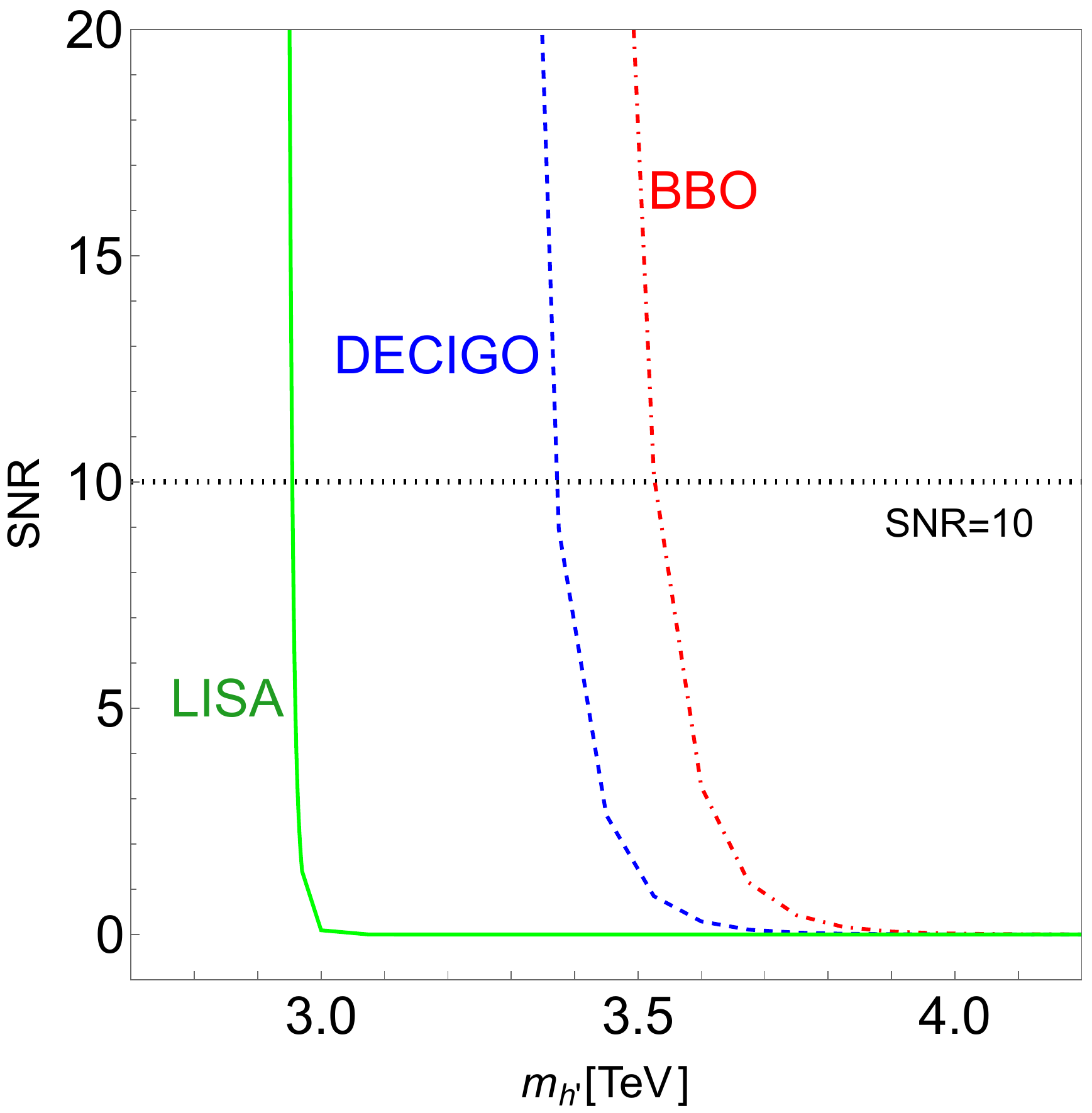}
\caption{
The value of SNR for $m_V=$ 3~TeV as a function of $m_{h'}$.  
Other parameters are $m_{h_D}=1.2 m_V$, $v_b=0.3$, and $\mathcal{T}$ = 4~yrs. 
The $m_{Z'}$ is adapted to explain the DM energy $\Omega h^2=0.12$. 
The green solid, blue dashed, and red dot-dashed lines represent the SNR values with $\mathcal{T}=4$ yrs and $v_b=0.3$ in the LISA, DECIGO, and BBO experiments, respectively.
The black horizontal dotted line represents SNR = 10. 
\label{fig:SNR5TeVDM}
}
\end{figure*}

Figure~\ref{fig:SNR5TeVDM} shows SNR as a function of $m_{h'}$ in the LISA, DECIGO, and BBO experiments. Here, we take $m_V=$ 3~TeV, $m_{h_D}=1.2 m_V$, $v_b=0.3$, and $\mathcal{T}$ = 4~yrs. We choose $m_{Z'}$ so as to explain the measured value of the DM energy density.
%
In the figure, the black horizontal dotted line represents SNR = 10, and each experiment can test the parameter region to the left of each intersection. 
We find that the LISA, DECIGO, and BBO experiments can detect the GW if
$m_{h'} < 2.95, 3.37,$ and 3.53~TeV, respectively.
The phase transition parameters depend on $\varphi_C/T_C$ as $\alpha\propto(\varphi_C/T_C)^2$ and $\beta/H\propto(\varphi_C/T_C)^{-5/2}$~\cite{Espinosa:2010hh,Caprini:2015zlo,Eichhorn:2020upj}. The large values of $\varphi_C/T_C$ are realized for smaller $m_{h'}$ as discussed in Sec.~\ref{subsec:FOFT}.
Therefore, the large values of SNR are obtained for small $m_{h'}$.

 \section{Results}
 \label{sec:numerical}

In this section, we explore the region of the parameter space where the GW detection experiments can test the model.
We also show the region that can be probed by the $W'$ search and can explain the measured value of the DM energy density via the freeze-out mechanism.
We focus on how the GW detection complement to the $W'$ search and study the following three cases for $m_{V^0}$:
\begin{itemize}
    \item For $m_{V^0} = 3$~TeV. This is the lightest mass of the vector DM in this model to obtain the right amount of the DM energy density via the freeze-out mechanism. In this case, the $W'$ searches already give the lower bound on $m_{Z'}$, and the HL-LHC can also test the model for $m_{Z'} \lesssim 8.5$~TeV~\cite{Abe:2020mph}.
    We discuss this case to see if the GW detection can probe the same region of the parameter space as the $W'$ search.

    \item For $m_{V^0} = 5$~TeV. In this case, there is no constraint from the current $W'$ search, but the $W'$ search in the HL-LHC can test the model for $m_{Z'} \lesssim 8.5$~TeV. We study the case for $m_{V^0} = 5$~TeV to investigate the complementary of the GW detection and the $W'$ search.

   \item For $m_{V^0} = 7$~TeV. The $W'$ search cannot test this case;  thus, we investigate this case how the GW detection is useful to test this model.   
\end{itemize}

\begin{figure*}[tb]
\centering
\includegraphics[width=0.4\textwidth]{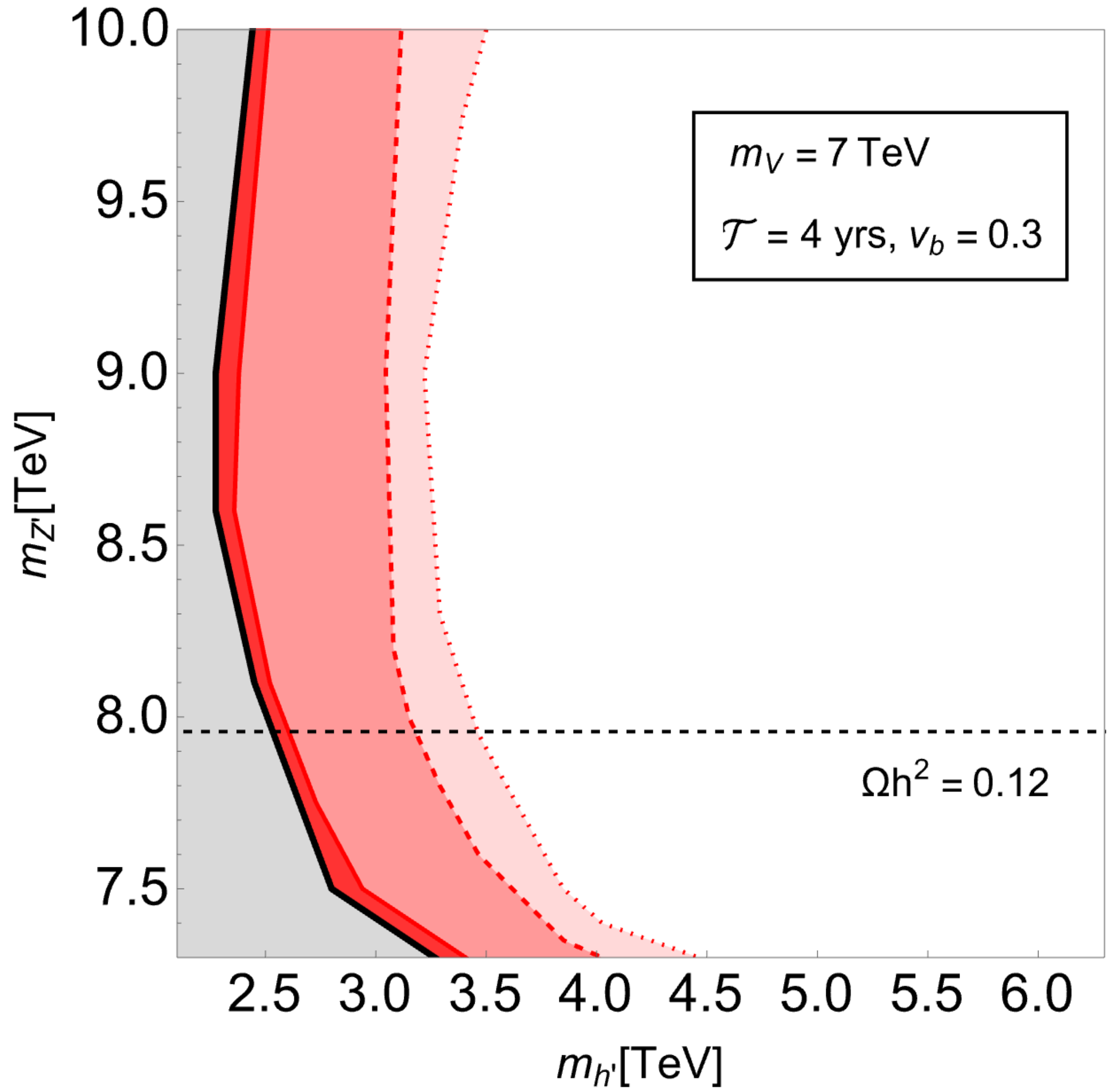}
\includegraphics[width=0.4\textwidth]{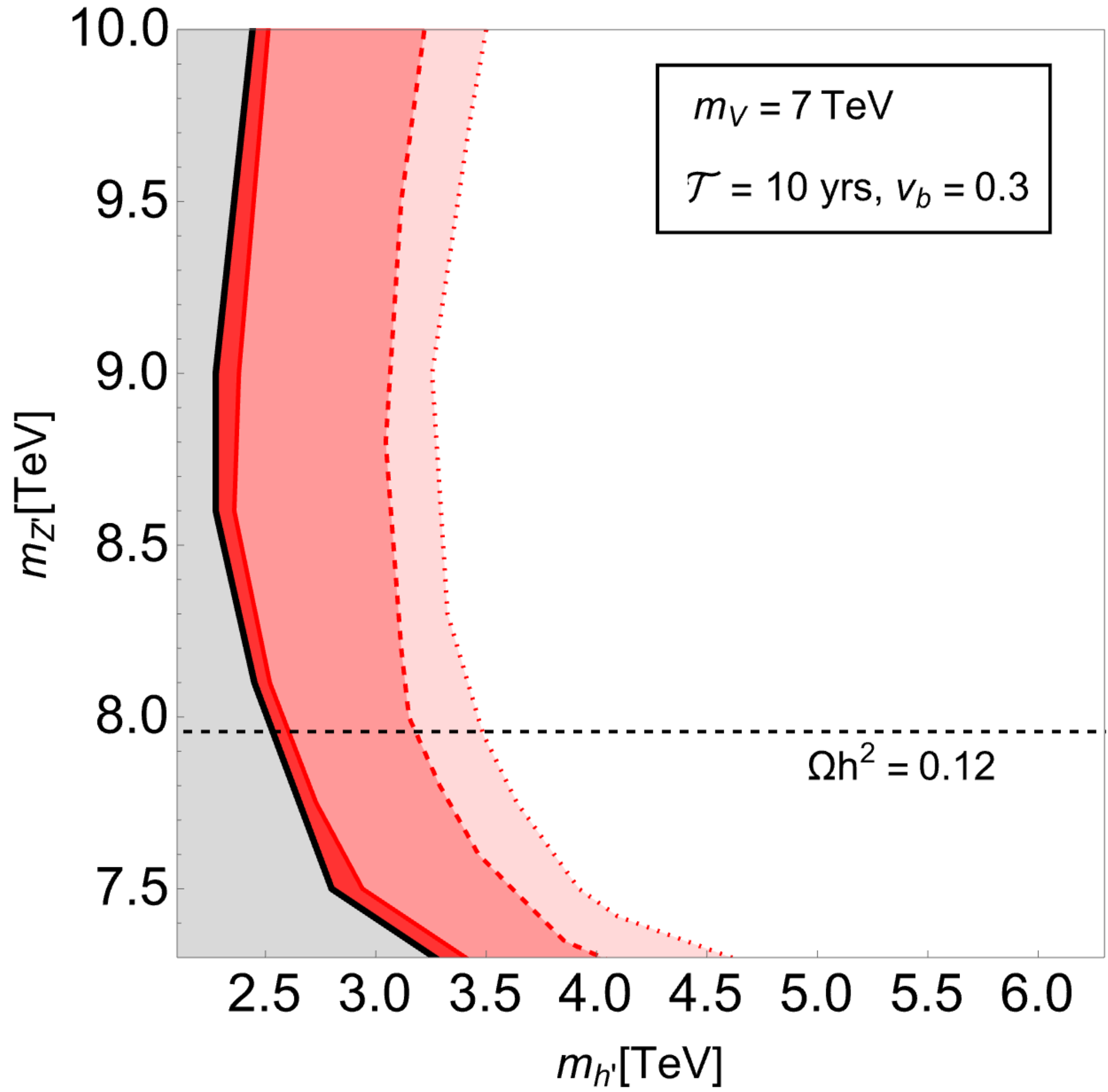}
\\[5mm] 
\includegraphics[width=0.4\textwidth]{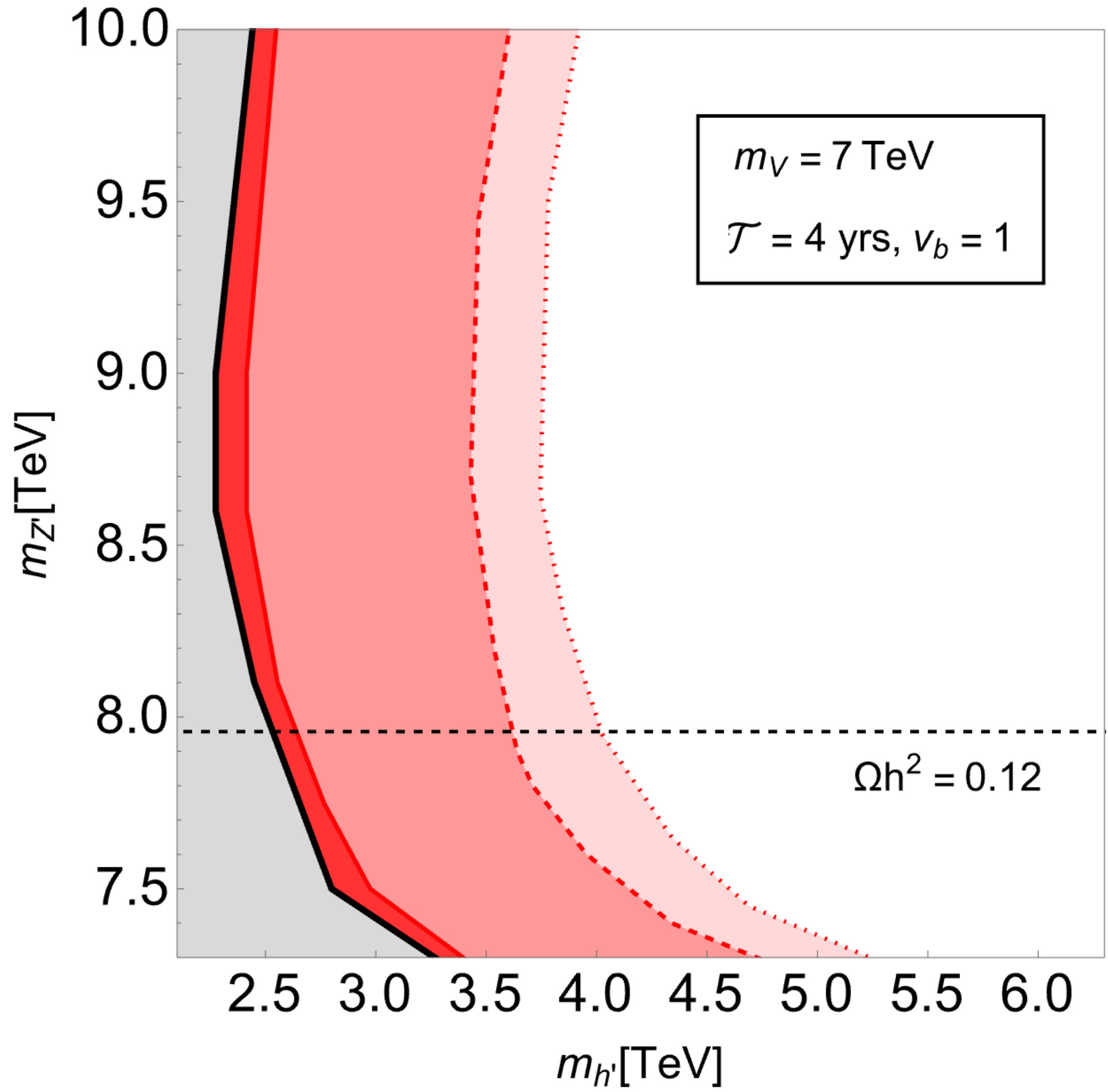}
\includegraphics[width=0.4\textwidth]{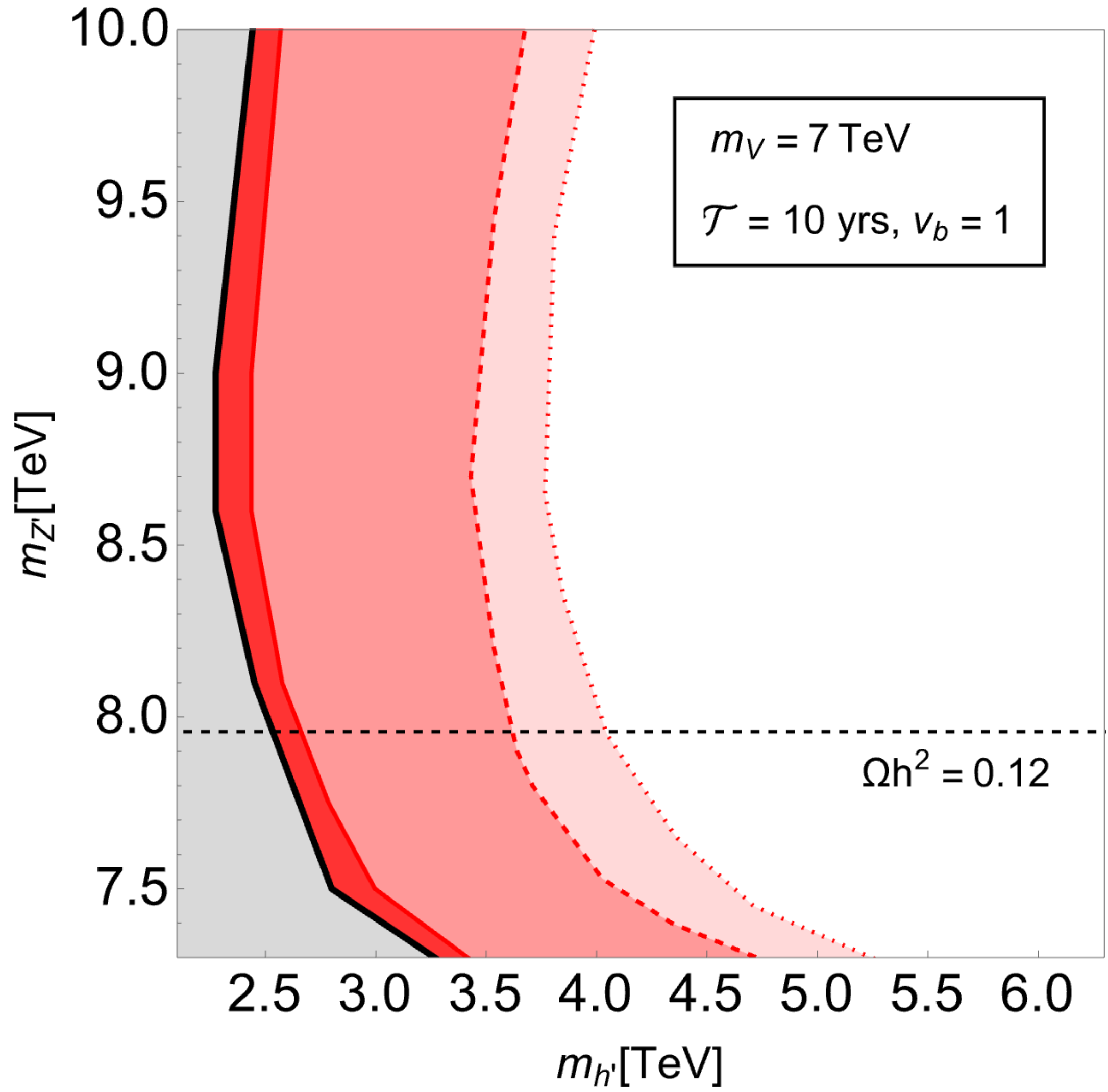}
\caption{
Detectability of the GW in the $m_{h'}$-$m_{Z'}$ plane for $m_V=$ 7~TeV and $m_{h_D}=1.2 m_V$. 
The upper (lower) two panels are for $v_b=$ 0.3 ($v_b =1$). 
In the left (right) panels, $\mathcal{T}$ = 4 (10)~yrs. 
In the light-red regions, SNR $>10$ in the BBO, DECIGO, and LISA experiments.
In the standard-red regions, SNR $>10$ in the DECIGO and LISA experiments.
In the dark-red regions, SNR $>10$ only in the LISA experiment.
The black-dashed lines indicate the regions where the measured value of the DM energy density is explained by the freeze-out mechanism. 
In the gray regions, which is to the left of the thick-black lines, the phase transition within the dark sector is not completed in the current universe. 
\label{fig:SNR7TeV}
}
\end{figure*}

First, we discuss the case for $m_{V^0} = 7$~TeV. In this case, the heavy vector and scalar bosons cannot be explored through the direct $W'$ search in the current and future collider experiments, and the importance of the GW detection increases. 
 Figure~\ref{fig:SNR7TeV} shows the region where SNR $>$ 10 in each GW experiment in the $m_{h'}$-$m_{Z'}$ plane for $m_V$ = 7 TeV and $m_{h_D}=1.2 m_V$. 
The upper (lower) two panels correspond to $v_b=0.3$ ($v_b = 0.1$).  
The colored regions can be tested by the GW detection experiments.
In the light-red region, the SNR is larger than ten in the BBO experiment. 
The standard-red (dark-red) regions 
 can be tested using DECIGO and BBO (LISA, DECIGO, and BBO) experiments. 
A strong first-order phase transition is not realized in the white regions to the right of the light-red region; thus, detectable GW spectra are not generated.
In the gray regions, 
the phase transition is not completed in the current universe, namely $\Gamma/H^4<1$. 
Along the black-dashed lines, the measured value of the DM energy density is explained by the freeze-out mechanism. 
We find that if $m_{Z'} \simeq 8$~TeV and 2.5~TeV $\lesssim m_{h'} \lesssim 3.5$~TeV, the model explains the measured value of the DM energy density and predicts the detectable GW simultaneously.
It is challenging to produce such a heavy $h'$ in collider experiments.
However, we can probe the heavy $h'$ regime using the GW signals.

\begin{figure*}[t]
\centering
\includegraphics[width=0.4\textwidth]{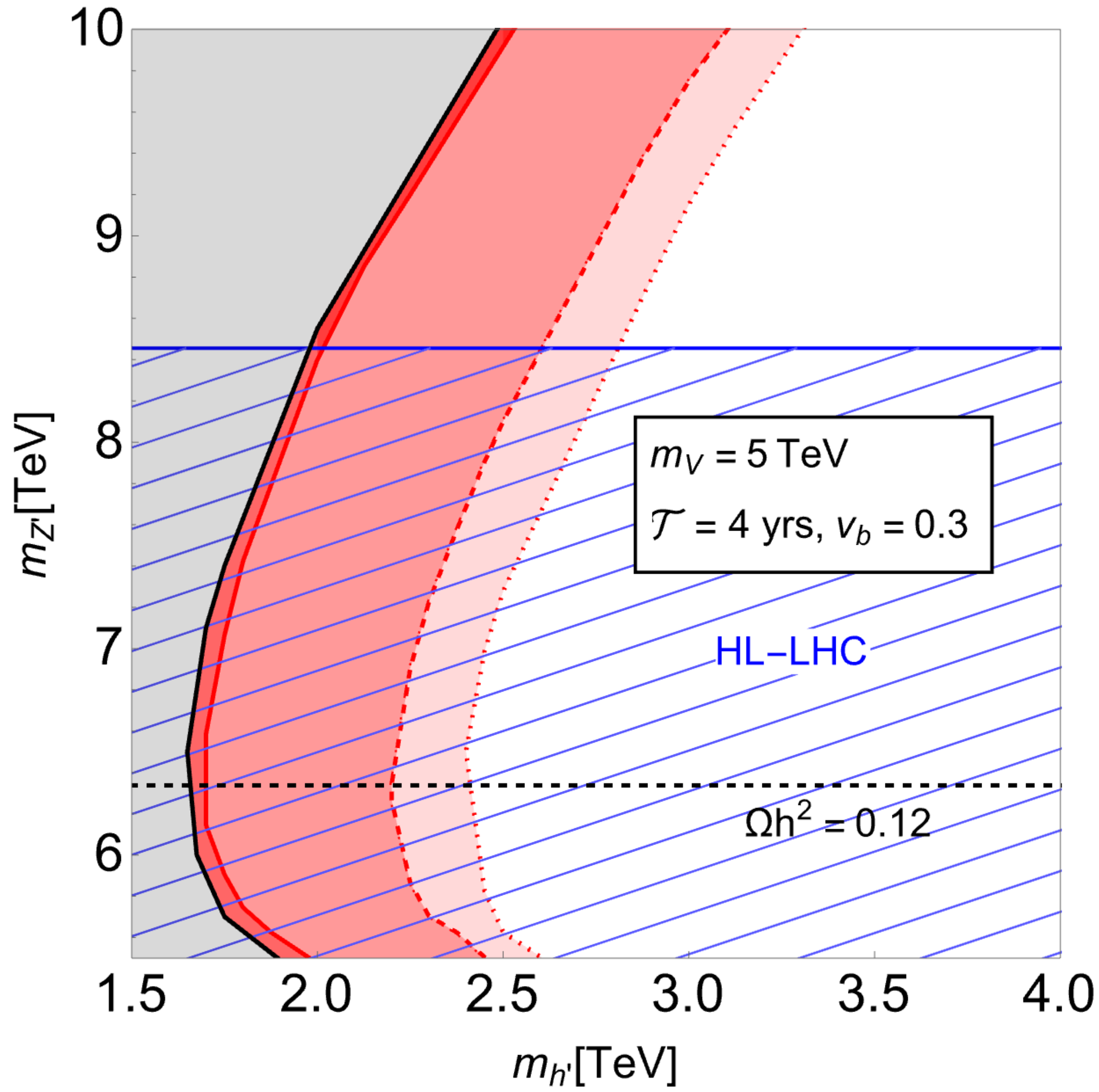}
\includegraphics[width=0.4\textwidth]{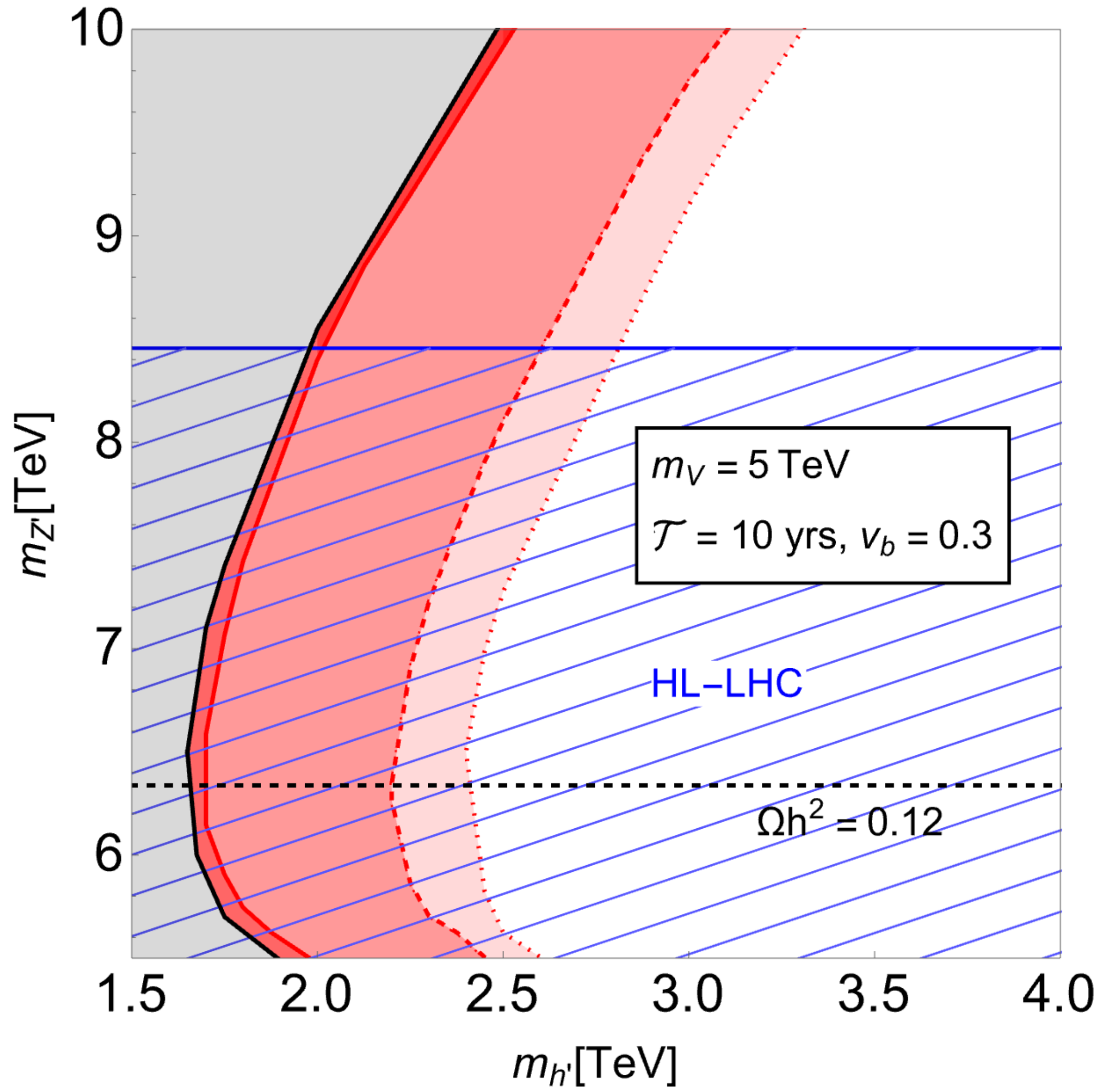}
\\[5mm] 
\includegraphics[width=0.4\textwidth]{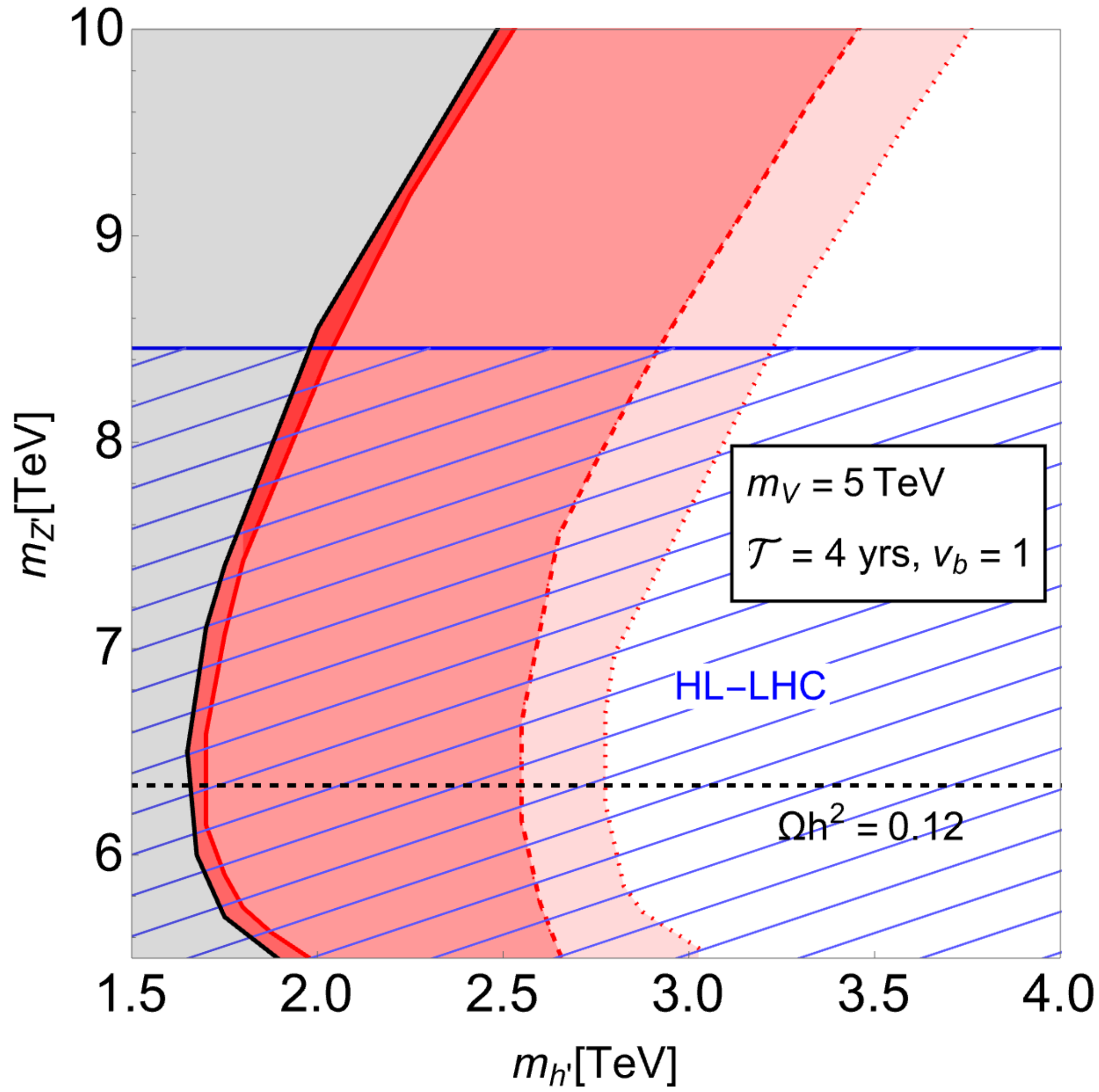}
\includegraphics[width=0.4\textwidth]{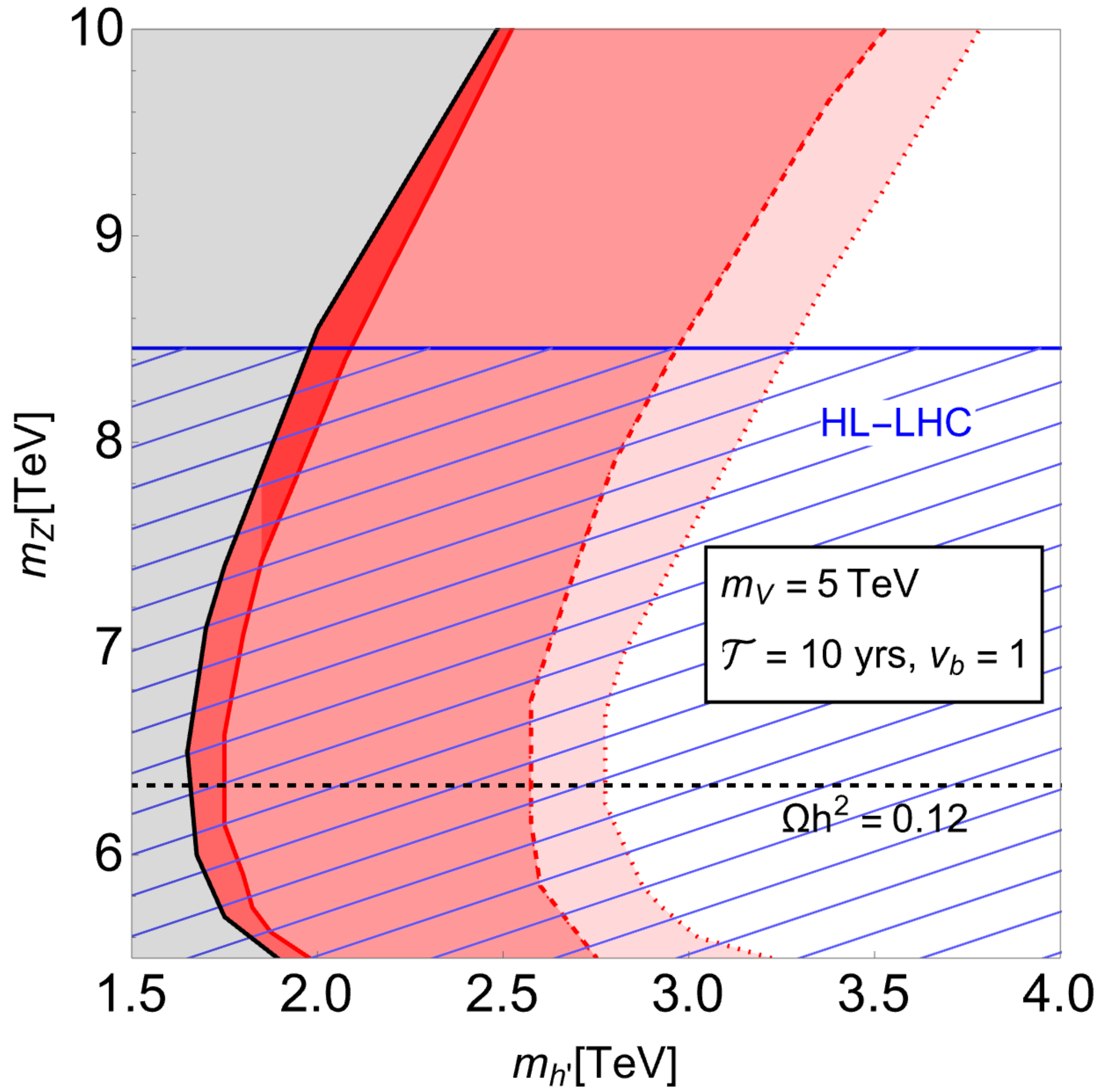}
\caption{
Detectability of the GW for $m_V=$ 5 TeV.
The blue-hatched regions are explored using the HL-LHC. 
The other color notations are the same as in Fig.~\ref{fig:SNR7TeV}.
\label{fig:SNR5TeV}
}
\end{figure*}

Next, we discuss the case for $m_V = 5$~TeV, where both the $W'$ search in the HL-LHC and the GW detection can be utilized to test the model.
Figure~\ref{fig:SNR5TeV} shows the result for $m_V$ = 5 TeV. 
The HL-LHC can discover $W'$ if the model parameters are within the blue-hatched regions.
The red-shaded regions can be probed using the GW. 
The black-dashed lines correspond to $\Omega h^2 = 0.12$.
Along the black-dashed lines, the blue-hatched and red-shaded regions are overlapped for $1.6\text{~TeV} \lesssim m_{h'} \lesssim 2.5$~TeV. 
Therefore, if we discover $W'$ at the HL-LHC and detect the GW, this mass range of $h'$ is the model prediction.
Because it is difficult to produce a heavy $h'$ in collider experiments, the GW signal is a useful tool to determine the range of $m_{h'}$.

\begin{figure*}[t]
\centering
\includegraphics[width=0.4\textwidth]{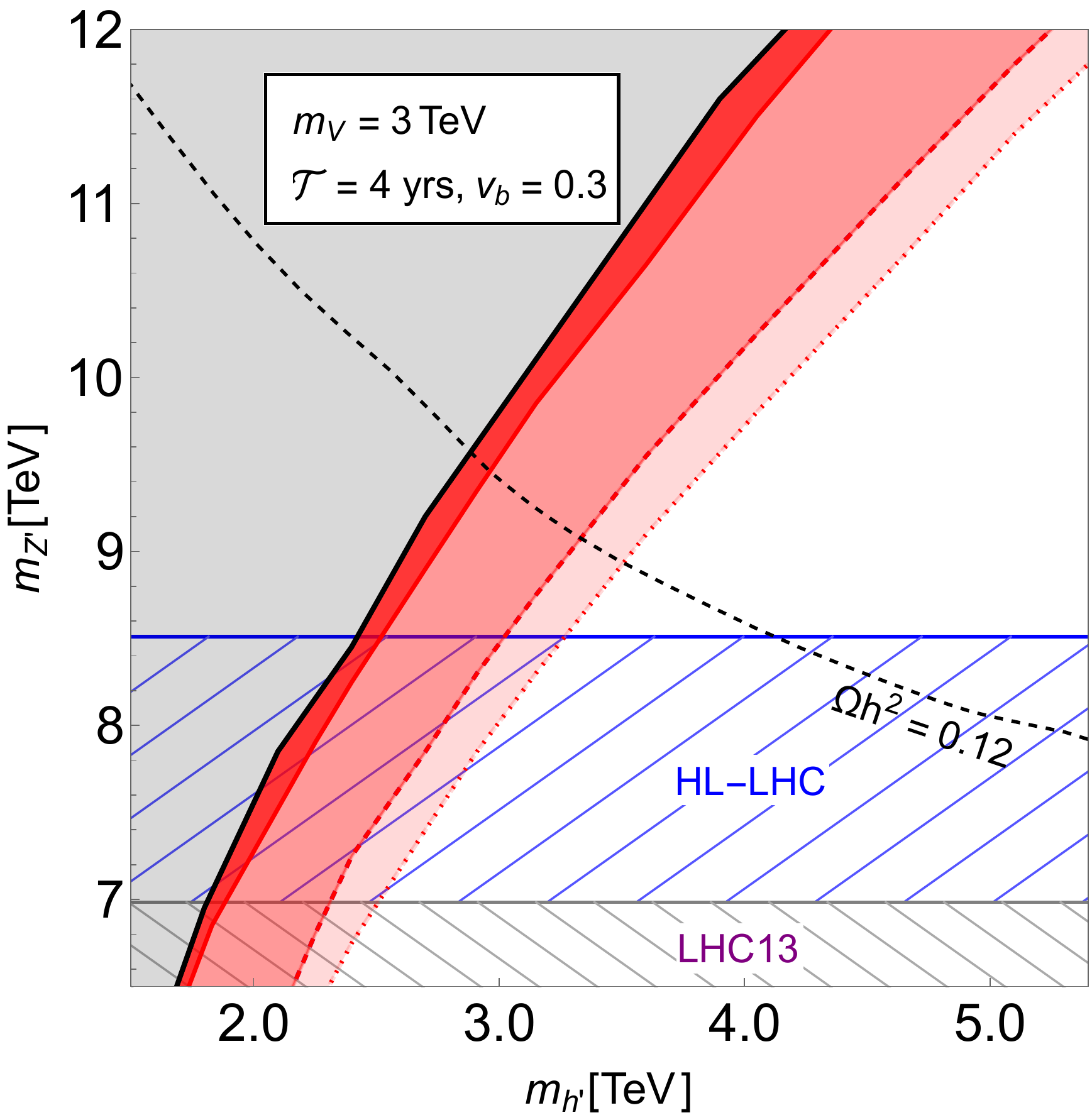}
\includegraphics[width=0.4\textwidth]{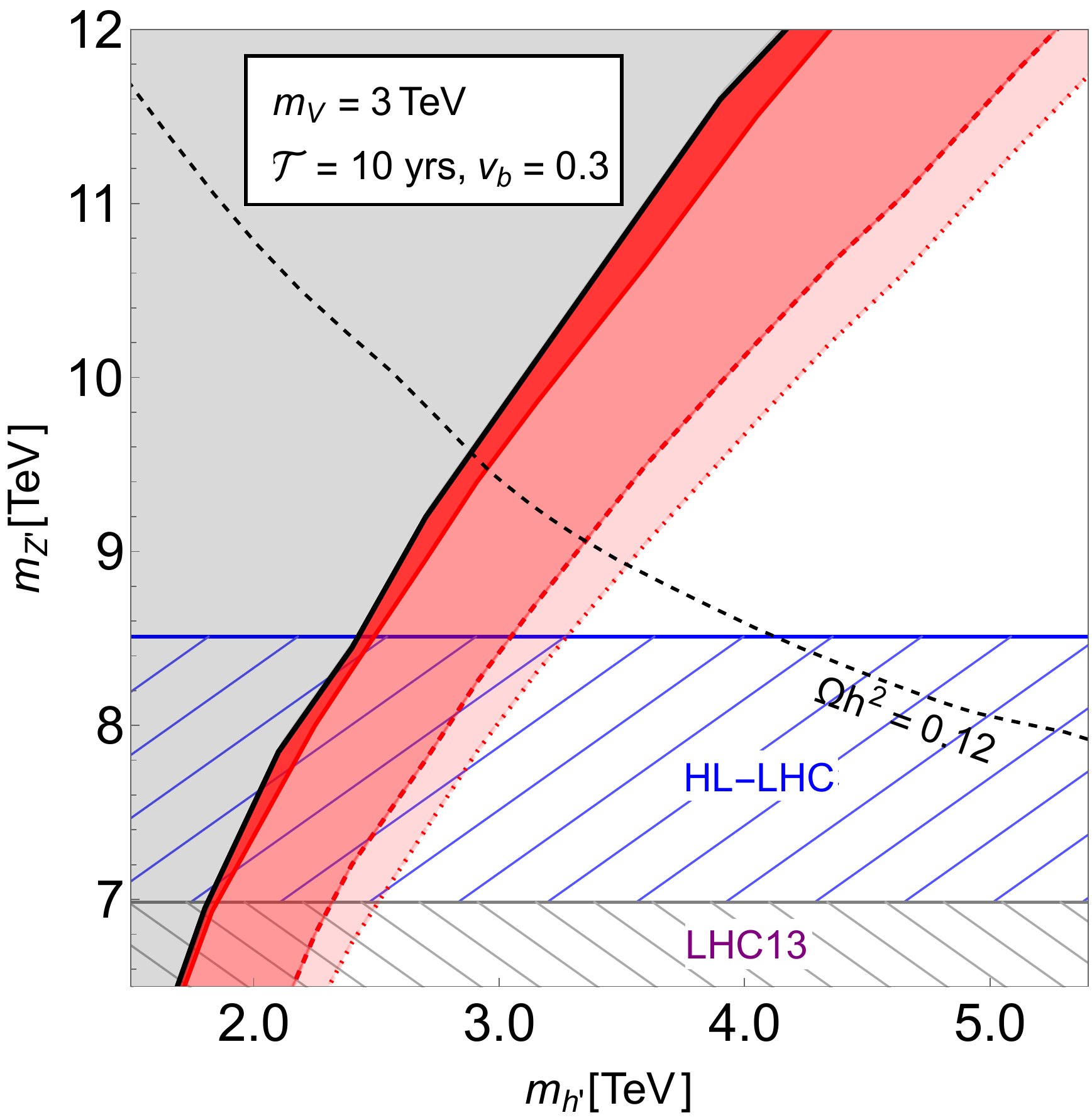}
\\[5mm] 
\includegraphics[width=0.4\textwidth]{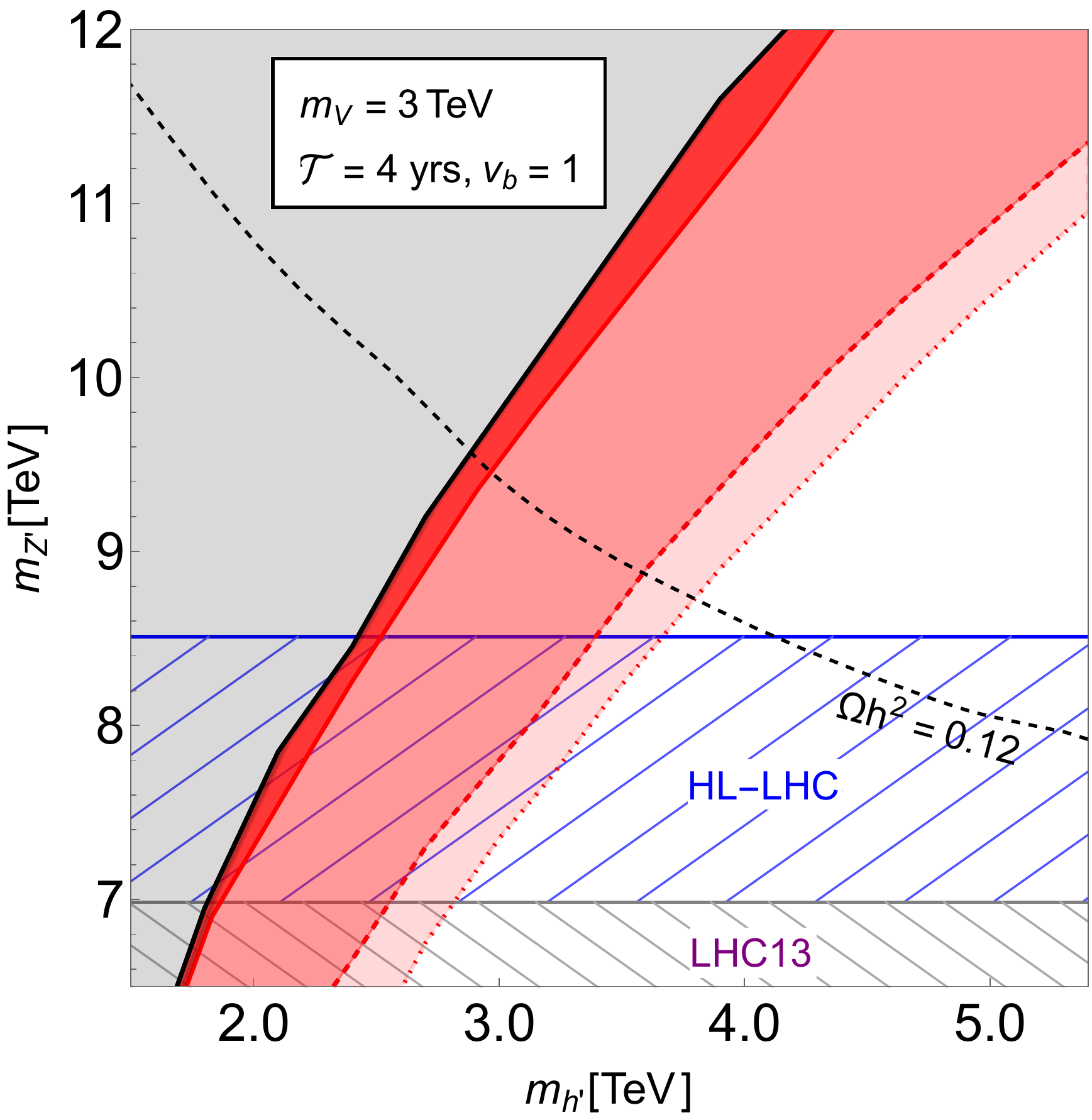}
\includegraphics[width=0.4\textwidth]{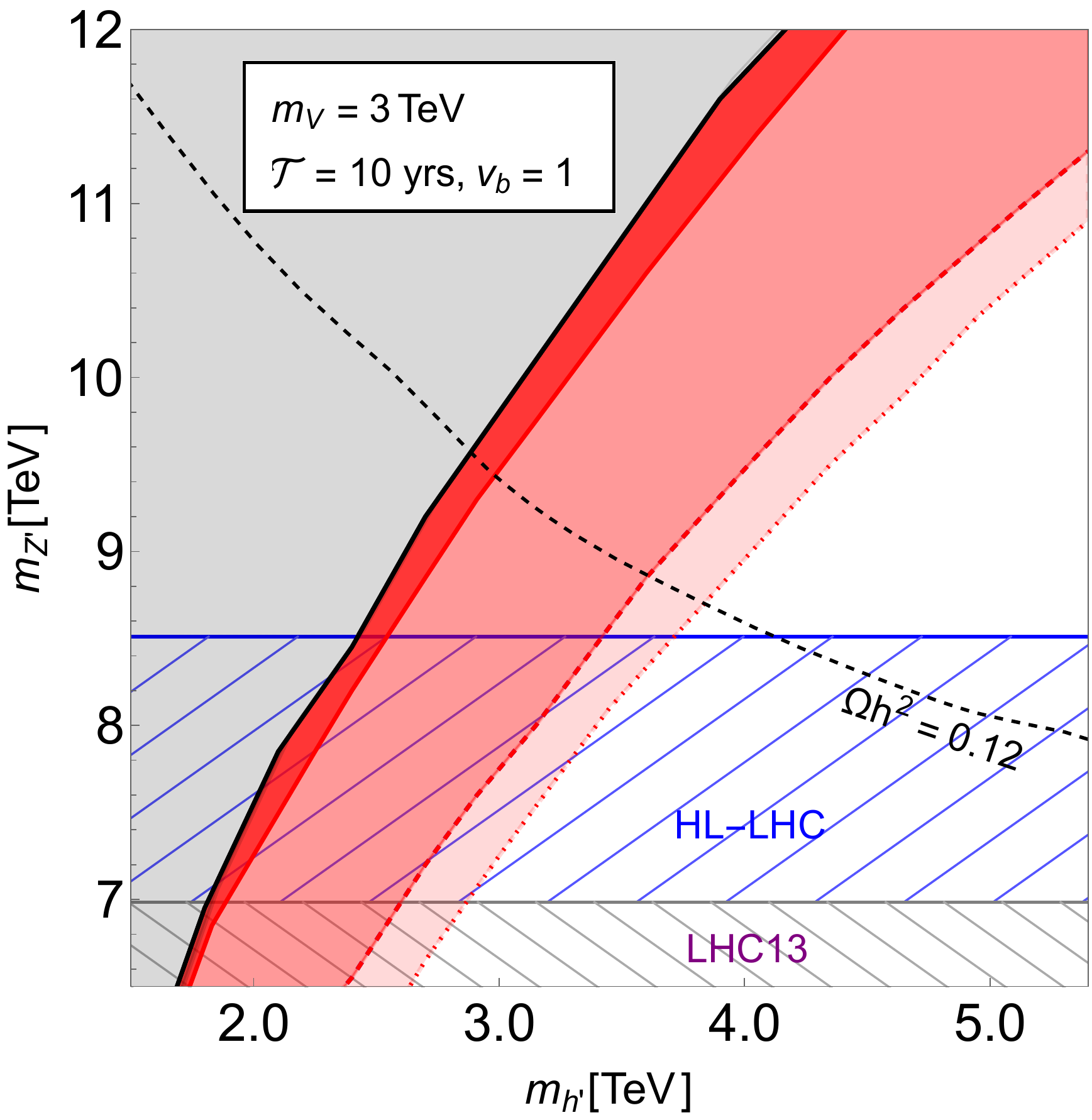}
\caption{
Detectability of the GW for $m_V=$ 3 TeV. 
The $W'$ search at the ATLAS experiment excludes the black-hatched region. 
The other color notation is the same as in Fig.~\ref{fig:SNR7TeV}.
}
\label{fig:SNR3TeV}
\end{figure*}

Finally, we discuss the case for $m_V = 3$~TeV.
The result is shown in Fig.~\ref{fig:SNR3TeV}. 
The direct search of $W'$ in the ATLAS experiment already excludes some regions of the parameter space.
For the parameter points that can explain the measured value of the DM energy density, we find that the $W'$ collider search and the GW observational experiments cover the different regions of the parameter space. 
The HL-LHC can probe the region for $m_{h'} \gtrsim 4$~TeV, and the GW observational experiments can probe the region for $2.8$~TeV $\lesssim m_{h'} \lesssim$ 3.5~TeV.
In this sense, the GW detection and the $W'$ search complement each other.

\clearpage
 \section{Conclusion}\label{sec:conclusion}

We have studied the GWs originating from the phase transition in the dark sector in the electroweakly interacting vector DM model
proposed in~\cite{Abe:2020mph}.

We have calculated the effective potential and investigated the phase transition. At the tree level, the potential has negative curvature at the origin. However, the gauge bosons give positive  contributions to the potential at the loop level, as shown in Eq.~\eqref{eq:Veff_T=0_around-origin}.
For the large gauge couplings, which is typically required to obtain the measured value of the DM energy density, the effective potential has positive curvature at the origin, even at $T=0$. 
As a result, the phase transition in the dark sector is first order and is strong, $\varphi_C/T_C \gtrsim 1$, in a wide range of the parameter space. 
The curvature at the tree level is proportional to $m_{h'}^2$; thus, the loop contributions are significant for smaller $m_{h'}$. Consequently, $\varphi_C/T_C$ is larger for smaller $m_{h'}$ as shown in Fig.~\ref{fig:VCTC}.
We also have found a lower bound on $m_{h'}$ for the phase transition. This is because the too small value of $m_{h'}$ makes the tunneling rate from the origin to the true vacuum too small, and then the phase transition does not occur.

We have studied three benchmarks ($m_V=7, 5$, and 3~TeV) and
found that the model predicts a GW spectrum that is detectable in the LISA, DECIGO, and BBO experiments.
Each benchmark has a different prediction for the $W'$ search in the collider experiments. The heavier $V^0$ cases cannot be tested by the $W'$ searches at the collider experiments, and thus, the GW detection is important to test the model. 
For $m_V = 7$ and 3~TeV, we have found that the GW detection can probe the region of the parameter space where the $W'$ searches cannot.
For $m_V = 5$~TeV, the region of the parameter space that the GW detection can probe overlaps with the region accessible by the $W'$ searches. However, the former region is narrower, and thus the GW is helpful in specifying the model parameters.
Assuming the model explains the measured value of the DM energy density via the freeze-out mechanism, we have found that the model predicts the detectable GW signals if $m_{h'}$ is a few TeV. Because it is challenging to search heavy $h'$ in collider experiments, utilizing the GW signals in determining $m_{h'}$ is crucial.

\section*{Acknowledgment}

This work was supported by JSPS KAKENHI Grant Number 19H04615 and 21K03549 [T.A.]. 
We would like to thank Editage (www.editage.jp) for English language editing.

\bibliographystyle{ptephy}
\bibliography{refs} 
 

\let\doi\relax



\appendix

\section{$V_\text{CW} + \delta V$}\label{app:Veff-at-T=0}
Defining $\varphi = \sqrt{\varphi_1^2 + \varphi_2^2}$, 
then the renormalized effective potential at $T=0$ as a function of $\varphi$ can be expressed as follows: 
\begin{align}
 \eval{V_\text{eff}}_{T=0}
=&
  \frac{m_{h'}^2-\Delta \Sigma_{h'h'}}{16 v_\Phi^2} 
  \left( (\varphi^2 - 2 v_\Phi^2)^2 - 4 v_\Phi^4 \right)
\nonumber\\
&
+ 9 \frac{m_V^4}{32\pi^2} \frac{\varphi^2}{2 v_\Phi^2}
+ 9 \frac{m_{Z'}^4}{32\pi^2} \frac{\varphi^2}{2 v_\Phi^2}
\nonumber\\
&
+ 9 \frac{m_V^4}{64\pi^2} \frac{\varphi^4}{4 v_\Phi^4}
\left( \ln \frac{\varphi^2}{2 v_\Phi^2} - \frac{3}{2} \right)
+ 9 \frac{m_{Z'}^4}{64\pi^2} \frac{\varphi^4}{4 v_\Phi^4}
\left( \ln \frac{\varphi^2}{2 v_\Phi^2} - \frac{3}{2} \right)
\nonumber\\
&
+
 \frac{1}{64\pi^2} \expval*{m_{h'}^2}^2 
\left(
\ln \frac{\expval*{{m}_{h'}^2}}{m_{h'}^2}
- \frac{3}{2}
\right)
+ 
\frac{1}{32\pi^2} 
\expval{{m}_{h'}^2} m_{h'}^2
\nonumber\\
&
+
 \frac{1}{64\pi^2} \expval{{m}_{h_D}^2}^2 
\left(
\ln \frac{\expval{{m}_{h_D}^2}}{m_{h_D}^2}
- \frac{3}{2}
\right)
+ 
\frac{1}{32\pi^2}
 \expval{{m}_{h_D}^2} m_{h_D}^2
\nonumber\\
&
- 6 \frac{(\lambda_{12}+ 2\lambda_\Phi)^2}{128\pi^2} 3 \varphi^2 (-4 v_\Phi^2 + \varphi^2)
\nonumber\\
&
+ 6 \frac{(\lambda_{12}+ 2\lambda_\Phi)^2 (\varphi^2 - 2 v_\Phi^2)^2}{64\pi^2}
\ln \frac{(\lambda_{12}+ 2\lambda_\Phi)(\varphi^2 - 2 v_\Phi^2)}{m_G^2}
\nonumber\\
&
+ \text{($\varphi$-independent terms)}
,
\end{align}
where  
\begin{align}
m_G^2 =& \lim_{\varphi \to \sqrt{2} v_\Phi} (\varphi^2 - 2 v_\Phi^2),\\ 
m_\Phi^2 =& -v_\Phi^2 (\lambda_{12}+2\lambda_\Phi)
,\\
\expval{m_{h'}^2} 
=&
-2 v_\Phi^2 (\lambda_{12}+2\lambda_\Phi)
+ (\lambda_{12} - 2 \sqrt{\lambda_{12}^2} + 6 \lambda_\Phi) \varphi^2
,\\
\expval{m_{h_D}^2} 
=&
-2 v_\Phi^2 (\lambda_{12}+2\lambda_\Phi)
+ (\lambda_{12} + 2 \sqrt{\lambda_{12}^2} + 6 \lambda_\Phi) \varphi^2
,\\
 \Delta \Sigma_{h'h'}
=& 
\frac{3 m_{h'}^2}{32\pi^2 v_\Phi^2}
\left( A_0(m_V^2) + A_0(m_{Z'}^2) \right)
\nonumber\\
&
+ \frac{3 m_{\Phi}^4}{2\pi^2 v_\Phi^2} B_0(0,0,0)
+ \frac{(\lambda_{12}-6\lambda_\Phi)^2 v_\Phi^2}{4\pi^2} B_0(0,m_{h_D}^2, m_{h_D}^2)
\nonumber\\
&
+ \frac{9 m_\Phi^4}{4\pi^2 v_\Phi^2} B_0(0,m_{h'}^2, m_{h'}^2)
+ \frac{9 m_V^4}{16\pi^2 v_\Phi^2} B_0(0,m_V^2, m_V^2)
\nonumber\\
&
+ \frac{9 m_{Z'}^4}{16\pi^2 v_\Phi^2} B_0(0,m_{Z'}^2, m_{Z'}^2)
\nonumber\\
&
- \frac{3(-m_{h'}^4 + 16 m_\Phi^4)}{32 \pi^2 v_\Phi^2} B_0(m_{h'}^2,0,0)
- \frac{(\lambda_{12}-6\lambda_\Phi)^2 v_\Phi^2}{4\pi^2} B_0(m_{h'}^2,m_{h_D}^2, m_{h_D}^2)
\nonumber\\
&
- \frac{9 m_\Phi^4}{4 \pi^2 v_\Phi^2} B_0(m_{h'}^2,m_{h'}^2,m_{h'}^2)
\nonumber\\
&
- \frac{3(m_{h'}^4 - 4 m_{h'}^2 m_V^2 + 12 m_V^4)}{64 \pi^2 v_\Phi^2} B_0(m_{h'}^2,m_V^2,m_V^2)
\nonumber\\
&
- \frac{3(m_{h'}^4 - 4 m_{h'}^2 m_{Z'}^2 + 12 m_{Z'}^4)}{64 \pi^2 v_\Phi^2} B_0(m_{h'}^2,m_{Z'}^2,m_{Z'}^2)
\nonumber\\
&
-m_{h'}^2 \delta_Z. 
\end{align}
Here, 
\begin{align}
A_0(m^2)
=& \frac{(4\pi)^2}{i} \int \frac{\dd[d]{\ell}}{(2\pi)^d} \frac{1}{\ell-m_1^2},
\\
B_0(p^2, m_1^2, m_2^2)
=& \frac{(4\pi)^2}{i} \int \frac{\dd[d]{\ell}}{(2\pi)^d} \frac{1}{(\ell-m_1^2)((\ell + p)^2 - m_2^2)},
\end{align}
and $\delta_Z$ is the counter term for the wave-function renormalization. 
 We choose the MS-bar renormalization condition for $\delta_Z$.  The IR divergences originated from the would-be NG boson contributions in $V_\text{CW}$ and $\Delta \Sigma_{h'h'}$ cancel each other. 
After the cancelation, the dominant contribution of $\Delta \Sigma_{h'h'}$ to the potential comes from the terms depending on the gauge couplings, described as 
\begin{align}
 \Delta \Sigma_{h'h'}
\simeq&
\frac{9 m_{h'}^2}{32\pi^2 v_\Phi^2} 
\left( m_V^2 \ln \frac{\mu^2}{m_V^2} + m_{Z'}^2 \ln \frac{\mu^2}{m_{Z'}^2}  \right)
.
\end{align}

\end{document}